\documentclass[aps,pra,twocolumn,superscriptaddress,showpacs]{revtex4-1}
\usepackage{graphicx,bm,color,float,textcomp,mathtools,times,nicefrac,soul,xcolor}
\newcommand\sho{$\rm SrHo_2O_4$}
\newcommand\sgo{$\rm SrGd_2O_4$}
\newcommand\seo{$\rm SrEr_2O_4$}
\newcommand\sdo{$\rm SrDy_2O_4$}
\newcommand\syo{$\rm SrYb_2O_4$}

\newcommand\slo{Sr$Ln_2$O$_4$}
\newcommand\alo{$ALn_2{\rm O}_4$}

\newcommand\gdi{Gd$^{3+}$}
\newcommand\afm{antiferromagnetic}
\newcommand\Tnone{$T_{\rm N1}$}
\newcommand\Tntwo{$T_{\rm N2}$}

\setstcolor{magenta}

\begin{document}
	\title{Magnetic structures of geometrically frustrated SrGd$_2$O$_4$ derived from powder and single-crystal neutron diffraction}
	\date{\today}
	\author{N.~Qureshi}		\affiliation{Institut Laue-Langevin, 71 Avenue des Martyrs, CS 20156, 38042 Grenoble Cedex 9, France}
	\author{B.~Z.~Malkin}	\affiliation{Kazan Federal University, Kazan, 420008, Kremlevskaya 18, Russia}
	\author{S.X.M.~Riberolles}\affiliation{Department of Physics, University of Warwick, Coventry CV4 7AL, United Kingdom}
	\author{C.~Ritter}		\affiliation{Institut Laue-Langevin, 71 Avenue des Martyrs, CS 20156, 38042 Grenoble Cedex 9, France}
	\author{B.~Ouladdiaf}	\affiliation{Institut Laue-Langevin, 71 Avenue des Martyrs, CS 20156, 38042 Grenoble Cedex 9, France}
	\author{G.~Balakrishnan}	\affiliation{Department of Physics, University of Warwick, Coventry CV4 7AL, United Kingdom}
	\author{M.~Ciomaga Hatnean}	\affiliation{Department of Physics, University of Warwick, Coventry CV4 7AL, United Kingdom}
	\author{O.~A.~Petrenko}	\affiliation{Department of Physics, University of Warwick, Coventry CV4 7AL, United Kingdom}	
\begin{abstract}	
We present the low-temperature magnetic structures of \sgo\ combining neutron diffraction methods on polycrystalline and single-crystal samples containing the $^{160}$Gd isotope.
In contrast to other members of the \slo\ family ($Ln$ = lanthanide) this system reveals two long-range ordered magnetic phases, which our diffraction data unambiguously identify.
Below \Tnone\ = 2.73~K, a $\mathbf{q}_1$ = (0 0 0) magnetic structure is stabilized where ferromagnetic chains along the $c$~axis (space group $Pnam$) are coupled antiferromagnetically with neighboring chains.
On cooling below \Tntwo\ = 0.48~K, an additional incommensurate component modulated by $\mathbf{q}_2$ = (0~0~0.42) evolves and aligned along either of the perpendicular axes for the two different Gd sites, resulting in a fan-like magnetic structure.
The identification of the particular Gd sites with the magnetic order observed with neutron diffraction is facilitated by a detailed analysis of the crystal fields acting on the sites.
The observed ordering phenomena underline the complex multiaxial anisotropy in this system.

\end{abstract}

\maketitle
\section{Introduction}
\begin{figure}[tb]
\includegraphics[width=0.78\columnwidth]{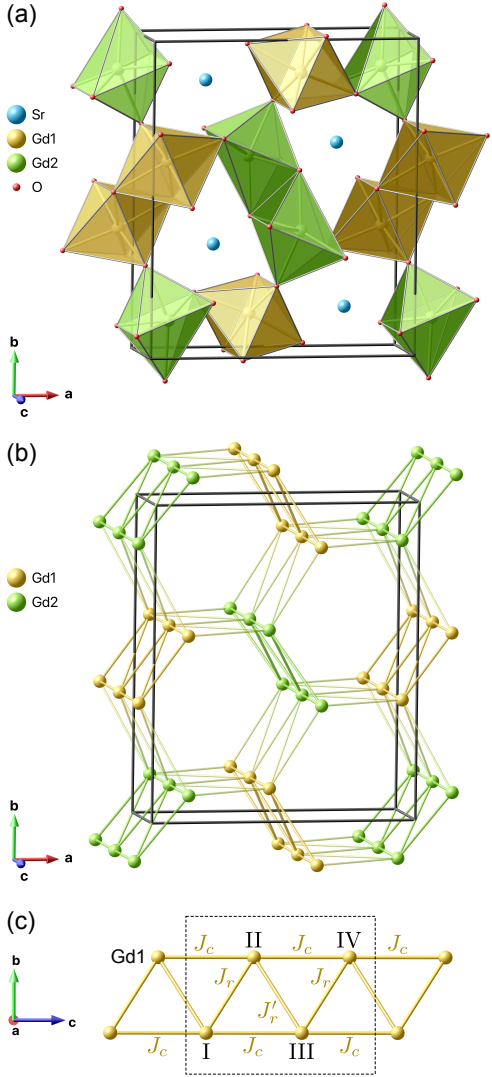}
\caption{Visualization of the crystal structure of \sgo.
(a) The oxygen octahedra around the Gd ions from the same site are edge-sharing, while the octahedra surrounding Gd ions from different sites are corner-sharing.
(b) The magnetic Gd ions form a distorted honeycomb lattice when projected onto the $ab$~plane and zig-zag ladders along the $c$~axis.
Nearest neighbors are connected by unicolor bonds along the $c$~axis for the Gd1 (yellow) and Gd2 sites (green), respectively.
The remaining unicolor bonds, the rungs of the zig-zag ladders, depict next-nearest neighbor links.
Further distances between the zig-zag ladders are shown by the bonds with a color gradient.
(c) A schematic of the magnetic interactions between the Gd ions in one of the zigzag chains.
A four-particle cluster used to model the magnetic properties of \sgo\ is highlighted by the dashed box.}
\label{fig:structure}
\end{figure}

The family of rare-earth oxides with general formula \alo, where $A$ = Ba, Sr and $Ln$ is a lanthanide,~\cite{Karunadasa_2005} has been of significant interest within the magnetism community due to its low magnetic ordering temperatures or (partial) absence of magnetic order.
In its particular crystal structure (space group $Pnam$) -- isostructural to calcium ferrite~\cite{Decker_1957} -- all atoms occupy Wyckoff position $4c$.
The magnetic $Ln^{3+}$ ions that occupy two distinct sites in the \alo\ compounds form distorted hexagons in the $ab$~plane which are connected along the short $c$~axis resulting in zig-zag ladders of edge-sharing triangles in a honeycomb-like arrangement (the crystal structure of this family of compounds is shown in Fig.~\ref{fig:structure} for the case of \sgo). 

Within such zig-zag ladders built exclusively from the ions occupying either $Ln$1 or $Ln$2 sites, the nearest neighbor (NN) interactions are along the legs of the ladder, while the slightly longer bonds of the next-nearest neighbors (NNN) build the rungs of the ladder.
Further neighbor interactions link the ladders between the different $Ln$ sites.
This particular geometry, in combination with the \afm\ exchange interactions, results in a strong geometrical frustration which is manifested in the suppression of magnetic ordering down to very low temperatures, complex ground states, as well as rich behavior under the application of a magnetic field like e.g. the appearance of magnetization plateaux.
In fact, the members of the \slo\ family are reported to order at temperatures well below their corresponding Weiss temperatures e.g. \sgo~\cite{Young_2014}, \seo~\cite{Karunadasa_2005,Petrenko_2008} and \syo~\cite{Quintero_2012}, or to remain (at least partially) disordered down to the lowest temperatures e.g. \sdo~\cite{Petrenko_2017,Gauthier_2017a,Gauthier_2017b} and \sho~\cite{Ghosh_2011,Young_2013}.
The different degrees of structural distortion within the oxygen octahedra surrounding the two $Ln$ sites result in strongly different crystal field energies which therefore dictate the single-ion anisotropy of the system and add to the compounds' complexity.
Many compounds from the \slo\ family demonstrate highly anisotropic magnetic properties at low-temperature.
It is also common for the magnetic moments on the $Ln$1 and $Ln$2 sites to behave almost independently, as the intrinsic coupling between the sites is rather limited.
In some cases, a short-range order on one $Ln$ site coexists with a long-range order on the other~\cite{Hayes_2011,Young_2013}. 

The Gd variant plays a special role among the \alo\ family due to the spin-only nature of the magnetic moment ($L=0$).
This intrinsic property should in principle lead to a rather isotropic behavior.
In addition, the spin-only nature of \gdi\ minimizes the crystal-field effects that are strong in other \slo\ materials.
The large magnetic moments borne by the \gdi\ ions are likely to induce significant dipolar interactions.
As a matter of fact, \sgo\ differs from the rest of the variants by the presence of two magnetic phase transitions at \Tnone\ = 2.73~K and at \Tntwo\ = 0.48~K as observed by specific heat measurements~\cite{Young_2014}.

We report in this paper on the low-temperature magnetic structures of \sgo.
Investigating Gd compounds by means of neutron scattering is particularly difficult due to its extremely high absorption cross section.
In this study, we use powder and single-crystal samples prepared using isotopically enriched Gd with $>$98\% of the low-absorbing $^{160}$Gd isotope.
Our investigations reveal a multi-axial anisotropy of the low-temperature magnetic phase in which -- as in other \slo\ compounds -- the two magnetic sites exhibit very different behavior.
This behavior is unexpected for the free \gdi\ ions, but there is a natural explanation when the crystal fields acting on the \gdi\ ions within the \sgo\ lattice are considered.

\section{Experimental details} \label{sec:experimental}
Polycrystalline samples of \sgo\ were prepared by the solid state method, as described in Refs.~\cite{Karunadasa_2005,Young_2014}.
Stoichiometric quantities of high purity SrCO$_3$ and isotopically enriched $\rm ^{160}Gd_2O_3$ powders were mixed, ground and heated to 1350~$^\circ$C in air for 48 hours.
The resulting material was then isostatically pressed into rods (approximately 5~mm diameter and 30~mm long) and sintered at 1100~$^\circ$C in air for 24 hours.
The isotopically enriched $\rm ^{160}Gd_2O_3$ was obtained from Trace Sciences International with the isotopic composition shown in Table~\ref{tab:Gdisotopes}.
Based on the supplied chemical analysis we have calculated the average scattering length and absorption cross section to be $b$ = (0.91 - 0.02i)$ \times 10^{-12}$~cm and $\sigma_a = 796.6$~barns, respectively.
These values were used to calculate the linear absorption coefficients for the absorption corrections and for the refinements of all the diffraction data.

\sgo\ crystals were grown by the floating zone technique~\cite{Balakrishnan_2009} using a two-mirror halogen furnace (NEC SC1MDH-11020, Canon Machinery Incorporated).
The growth was carried out in an argon atmosphere at a pressure of 2~bars, using a growth speed of 3~mm/h.
The two rods (feed and seed) were counter-rotated at a rate of 15~rpm.
From a crack-free translucent crystal boule with a length of approximately 1~cm, seven single crystal fragments were extracted.
A sample of 0.0127~g that showed the best crystalline quality was chosen for the neutron diffraction experiments.

The phase purity of powder as well as quality of single-crystal samples were confirmed by means of powder X-ray and X-ray Laue diffraction, respectively.
Rietveld analysis revealed traces of monoclinic Gd$_2$O$_3$ resulting in a powder sample purity of 98.2\%.
The sequence of magnetic transitions was confirmed through magnetization and specific heat experiments. 

\begin{table}[tb]
	\caption{Isotopic composition of the Gd$_2$O$_3$ powder purchased from Trace Sciences International.
	An average scattering length of $b$ = (0.91 - 0.02i)$ \times 10^{-12}$~cm and absorption cross section of $\sigma_a = 796.6$~barns was used for all the refinements and corrections.
	The values for the scattering lengths and absorption cross sections were taken from~\cite{sea1992}.}
	\label{tab:Gdisotopes}
	\begin{ruledtabular}
		\begin{tabular}{cccc}
			Isotope   &  $b$ ($10^{-12}$ cm)  &  $\sigma_a$ (barns)  &  Enrichment (\%)   \\ \hline
			$^{152}$Gd & 1.0(3) & 735(20) & $<$0.06 \\
			$^{154}$Gd & 1.0(3) & 85(12) & 0.03 \\
			$^{155}$Gd & 0.6-1.7i & 61100(400) & 0.2 \\
			$^{156}$Gd & 0.63& 1.5(1.2) & 0.32 \\
			$^{157}$Gd & -0.114-7.19i & 259000(700) & 0.26 \\
			$^{158}$Gd & 0.9(2) & 2.2 & 0.79 \\
			$^{160}$Gd & 0.915 & 0.77 & 98.4(1) \\
		\end{tabular}
	\end{ruledtabular}
\end{table}

Powder neutron diffraction (PND) experiments were carried out on the high-resolution diffractometer D2B and on the high-flux diffractometer D20 (both ILL, Grenoble) using a wavelength of $\lambda_{\rm{D2B}} = 1.594$~\AA\ and $\lambda_{\rm{D20}} = 2.41$~\AA, respectively.
The high-resolution experiment was performed at room temperature, whereas the high-flux data collection was done in a standard orange cryostat for measurements down to 1.45~K as well as in a dilution fridge with a base temperature of 70~mK.
Due to the still non-negligible absorption of the isotopically enriched Gd the powder was packed into sachets of Cu foil in order to make use of the total height of the sample container made out of Cu and V for the dilution and standard cryostat experiments, respectively.
The thickness of the powder within the sample holder is estimated to be less than 1~mm. 

Single-crystal neutron diffraction experiments were performed on the D10 diffractometer (ILL, Grenoble) with a wavelength of \mbox{$\lambda_{\rm{D10}} = 2.36$~\AA} supplied by the (002) reflection of a pyrolytic graphite monochromator.
The instrument was equipped with a unique 4-circle dilution cryostat allowing a minimum temperature of 100~mK without being confined to a single scattering plane.
An appropriate absorption correction was carried out using \textsc{Mag2Pol} \cite{Mag2pol} by drawing a convex-hull model of the sample in order to calculate the mean beam path for every single Bragg reflection measured.
Irreducible representations were calculated and all single-crystal data were analyzed using \textsc{Mag2Pol}~\cite{Mag2pol}.

\section{Results and analysis}
\subsection{Powder neutron diffraction} \label{pnd}
\begin{table}[tb] 
\caption{Refined structural parameters of \sgo\ within the orthorhombic $Pnam$ space group obtained from the PND experiments performed on the instruments D2B at room temperature and on D20 at 4~K respectively.
	All ions are in the Wyckoff $4c$ positions with coordinates ($x$ $y$ $\nicefrac{1}{4}$). }
	\begin{ruledtabular}
		\begin{tabular}{ccccc}
			& \multicolumn{2}{c}{D2B $T$ = 300~K}	& \multicolumn{2}{c}{D20 $T$ = 4~K}	\\
			Atoms	&	$x$		&	$y$		&	$x$		&	$y$		\\	\hline
			Sr		&	0.7500(2)	&	0.6492(2)	&	0.7504(4)	&	0.6489(4)	\\
			Gd1		&	0.4263(2)	&	0.1131(1)	&	0.4263(4)	&	0.1135(3) 	\\
			Gd2		&	0.4171(2)	&	0.6112(1)	&	0.4187(4)	&	0.6116(2)	\\
			O1		&	0.7157(3)	&	0.3187(2)	&	0.7185(4)	&	0.3201(4)	\\
			O2		&	0.6312(3) 	&	0.0177(2)	&	0.6300(4)	&	0.0167(4)	\\
			O3		&	0.5077(2)	&	0.7845(2)	&	0.5099(5)	&	0.7855(3) \\
			O4		&	0.9266(3) 	&	0.0795(2)	&	0.9285(6)	&	0.0778(3)	\\	\hline
			$a$ (\AA)	& \multicolumn{2}{c}{10.1312(3)}		& \multicolumn{2}{c}{10.1000(3)}	\\ 
			$b$ (\AA)	& \multicolumn{2}{c}{12.0599(3)}		& \multicolumn{2}{c}{12.0395(4)}	\\ 
			$c$ (\AA)	& \multicolumn{2}{c}{3.47519(9)}		& \multicolumn{2}{c}{3.4682(1)}		\\ 
			$R_{\rm F}$      & \multicolumn{2}{c}{2.83}				& \multicolumn{2}{c}{6.68} 
		\end{tabular}
	\end{ruledtabular}
	\label{tab:nucpowder}
\end{table}

\begin{figure}[tb]
\includegraphics[width=0.9\columnwidth]{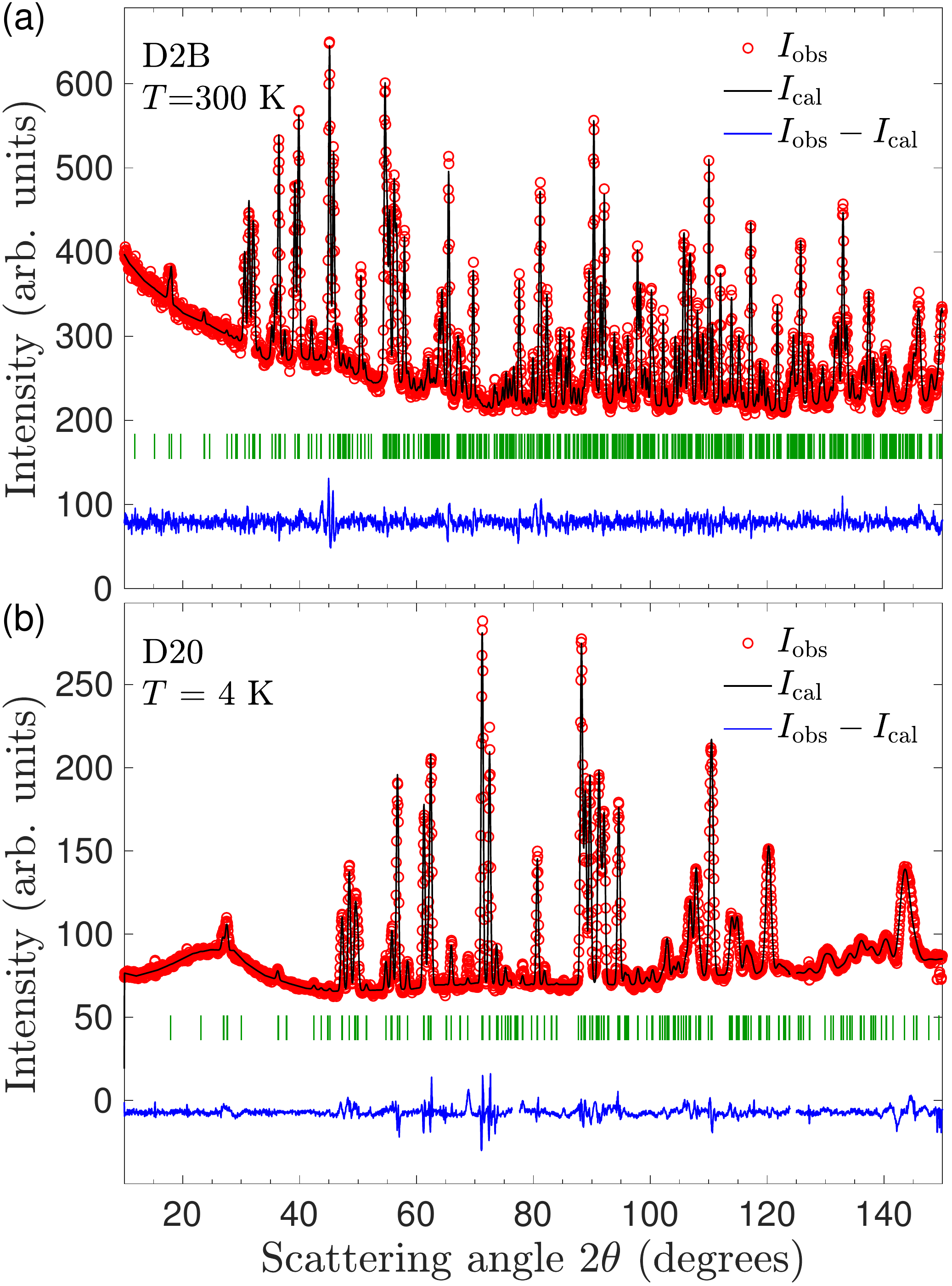}
\caption{Powder neutron diffraction pattern in the paramagnetic phase of \sgo.
The observed (red dots) and calculated (black line) patterns are shown together with the difference curve (blue line) and the markers of nuclear peak positions (green lines) for the (a) D2B measurement at room temperature and (b) the D20 measurement at \mbox{4 K}.
A strong diffuse scattering signal is seen in the paramagnetic regime at 4~K as a broad peak around 27~degrees.}
\label{fig:nucpattern}
\end{figure}

The isotopically enriched \sgo\ sample was measured on the high-resolution diffractometer D2B at room temperature, the results are shown in Fig.~\ref{fig:nucpattern}(a).
According to the description in Sec.~\ref{sec:experimental} a linear absorption coefficient of $\mu = 1.33$~mm$^{-1}$ was used with the employed wavelength of $\lambda_{\rm{D2B}} = 1.594$~\AA.
By varying the overall scale factor, the lattice parameters, the atomic positions and an isotropic temperature factor $B$, the best agreement factor together with a reasonable $B{\rm (Gd)}$ = 0.37(2)~\AA$^2$ was achieved for a sample thickness of 0.77~mm which agrees very well with the prepared powder in the Cu sachets.
A diffraction pattern within the paramagnetic phase was recorded at $T = 4$~K on the high-flux diffractometer D20 [Fig.~\ref{fig:nucpattern}(b)].
In this case, the linear absorption coefficients amounts to $\mu$ = 2.07 mm$^{-1}$ and the sample thickness was kept constant.
Due to the limited $Q$ range in comparison to the D2B pattern, an overall isotropic temperature factor was used which was refined to 0.25(6)~\AA$^2$.
The refined structural parameters from the experiments on both instruments are compared in Table~\ref{tab:nucpowder}.

\begin{figure}[tb]
\includegraphics[width=0.9\columnwidth]{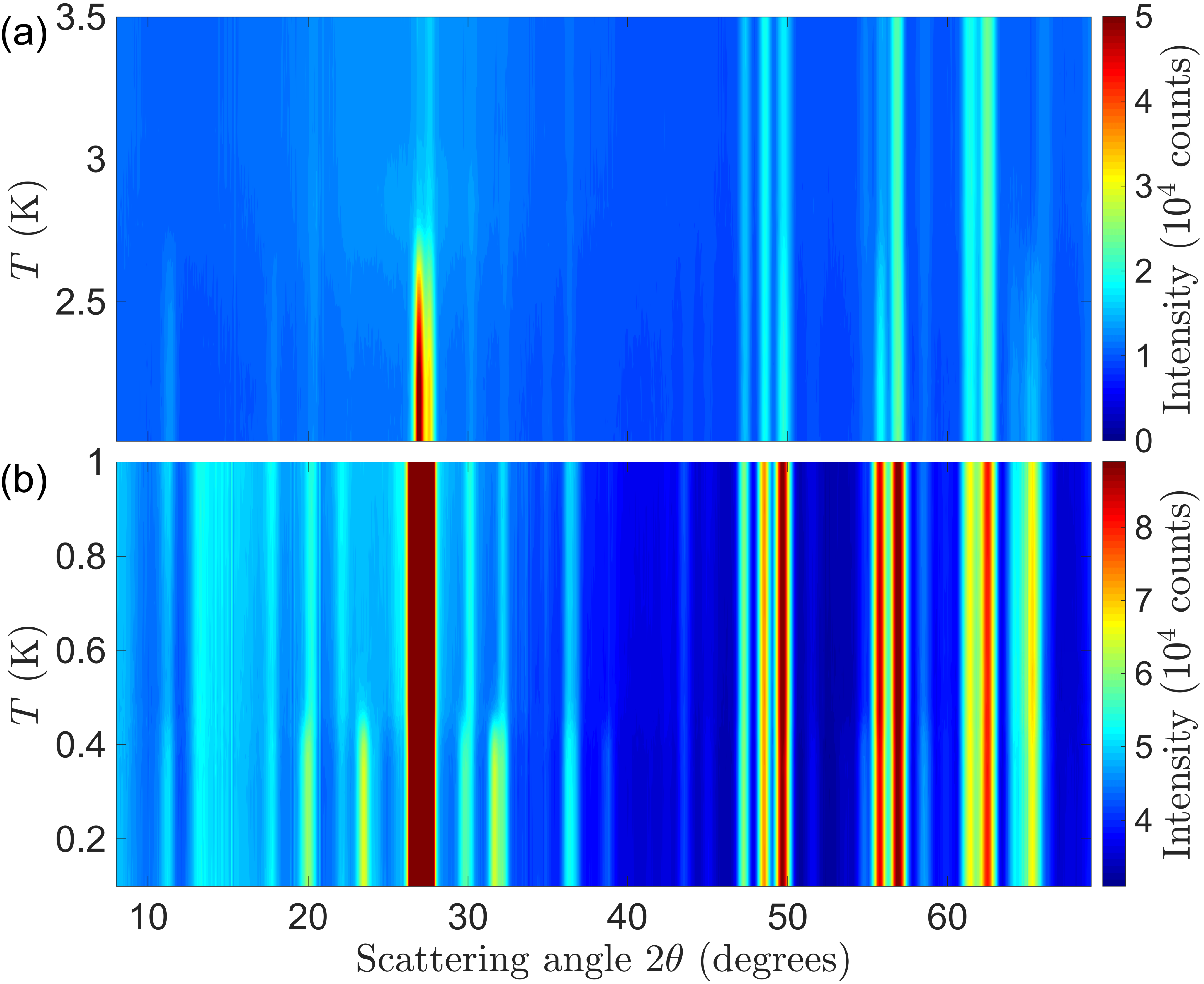}
\caption{Thermodiffractograms (D20, $\lambda$ = 2.41~\AA) revealing the magnetic phase transitions at (a) \Tnone\ = 2.73~K and (b) \Tntwo\ = 0.48~K.
In (a), the onset of the commensurate \afm\ structure is manifest by the $\mathbf{q}_1$ = (0 0 0) reflections while in (b), the satellites appear at positions different to those given by the translation symmetry of the underlying nuclear structure.}
\label{fig:magthermo}
\end{figure} 

Figure~\ref{fig:magthermo} shows two thermodiffractograms recorded using a standard cryostat (upper panel) and a dilution insert (lower panel) tracking the magnetic phase transitions at \Tnone\ and \Tntwo, respectively.
For the former, diffraction patterns have been recorded for 30 minutes with a temperature interval of roughly 0.15~K, while for the latter, a counting time of 1~hour was used with steps of 25~mK around \Tntwo\ and 100~mK between 0.6 and 1~K. 

\begin{figure}[tb]
\includegraphics[width=0.9\columnwidth]{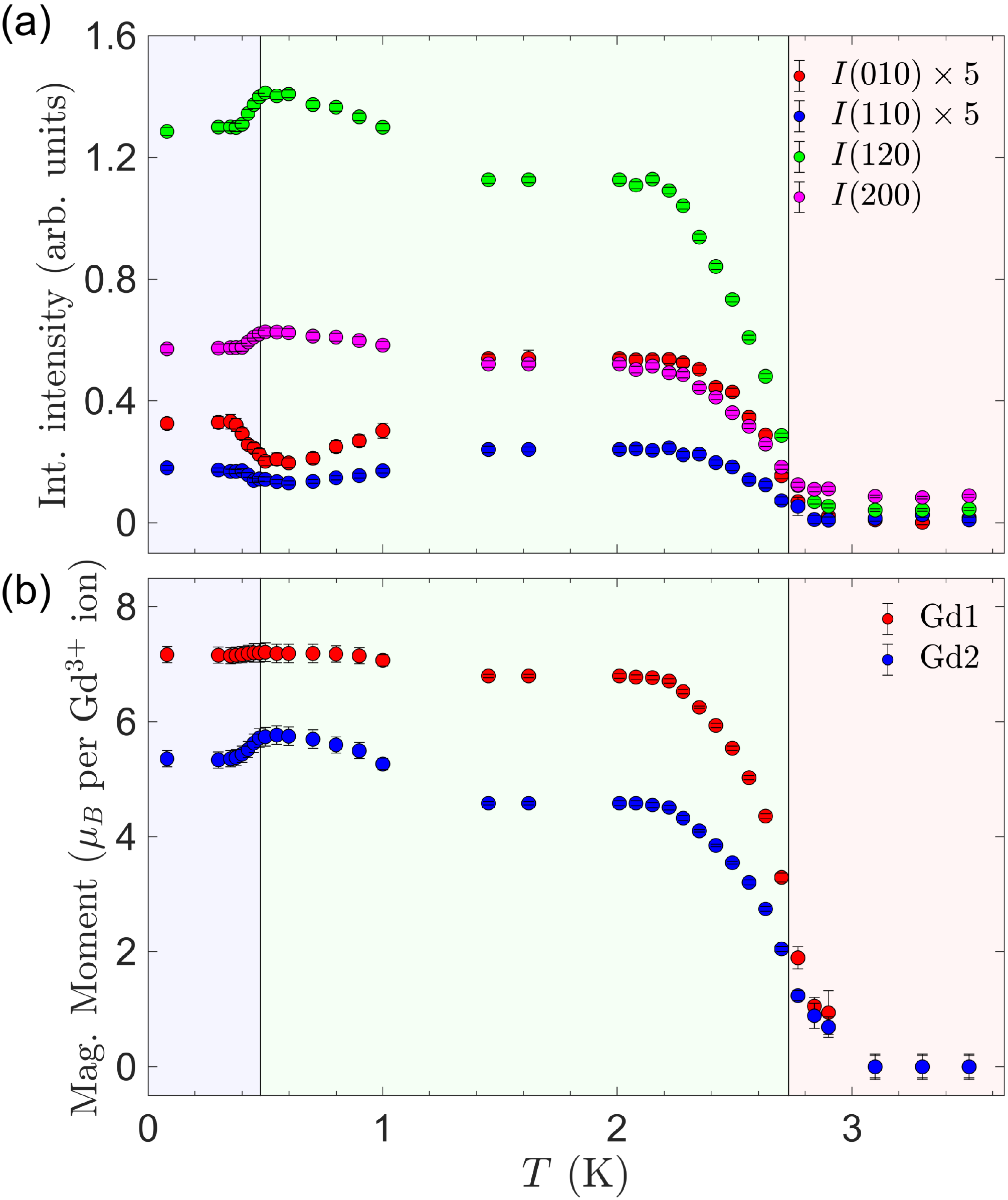}
\caption{(a) Integrated intensities as a function of temperature (data shown in Fig.~\ref{fig:magthermo}) for selected Bragg reflections with strong magnetic contribution normalized to the integrated intensity of a nuclear peak at large $Q$ with negligible magnetic contribution.
The 3 regimes \mbox{$T > $ \Tnone}, \mbox{\Tntwo~$< T <$ \Tnone} and \mbox{$T < $ \Tntwo} are emphasized by different background colors.
While the intensities of all 4 reflections increase monotonically with decreasing temperature above \Tnone, a more intricate behavior is present below \Tnone: the (120) and (200) reflections (at $2\theta$ values 27.0$^\circ$ and 27.6$^\circ$, respectively, in Fig.~\ref{fig:magthermo}) slowly increase in intensity before showing a sharp drop at \Tntwo\ whereas the (010) and (110) ($2\theta = 11.4^\circ$ and $2\theta = 17.9$, respectively) show exactly the opposite behaviour.
(b) Refined magnetic moments along the $c$ axis (taking into account the commensurate $\mathbf{q}_1$ reflections only) revealing a stronger increase of the Gd2 magnetic moment in the temperature range between \Tnone\ and \Tntwo\ compared to that of Gd1, but also a stronger decrease when entering the low-temperature phase at \Tntwo.}
\label{fig:tdep_pow}
\end{figure} 

In Fig.~\ref{fig:magthermo}(a), the appearance of a very intense magnetic peak at 2$\theta = 27.1^\circ$ accompanied by numerous weaker peaks (e.g. at $11.5^\circ$ or $18.0^\circ$ in $2\theta$) at around $T=2.7$~K marks the onset of the magnetically ordered state, which is in a very good agreement with the previously reported \Tnone\ = 2.73~K~\cite{Young_2014}.
In Fig.~\ref{fig:magthermo}(b), magnetic satellites can be observed below $T \sim 0.45$~K (e.g. at scattering angles $19.9^\circ$, $23.4^\circ$, $29.7^\circ$ or $31.6^\circ$) at positions which do not coincide with a commensurate structure.
This second transition temperature also coincides with \Tntwo\ = 0.48~K reported from the specific heat measurements~\cite{Young_2014}.
While the onset of the respective magnetic phases is clearly visible in the color intensity map, Fig.~\ref{fig:magthermo} falls short in showing the details of the temperature dependence.
Therefore, the Bragg reflections (010), (110), (120) and (200) with strong magnetic contribution at scattering angles 11.5$^\circ$, 18.0$^\circ$, 27.1$^\circ$ and 27.6$^\circ$, respectively, were integrated using a Gaussian profile on a sloping background for all temperatures.
The integrated intensities (normalized to a strong nuclear peak with negligible magnetic contribution at the scattering angle of 62$^\circ$) are shown in Fig.~\ref{fig:tdep_pow}(a).
The diffraction patterns recorded at $T = 1.45$~K in the standard orange cryostat and at $T=1.6$~K in the dilution fridge were used to bring the integrated intensities of the different instrument configurations to the same scale.
The 4 selected reflections reveal a monotonic increase in intensity when entering the magnetically ordered phase below \Tnone, until a saturation value is reached around 2~K.
Below \mbox{$T$ = 1.45 K} a more intricate evolution occurs, in which the (010) and (110) reflections decrease in intensity when approaching \Tntwo, while the intensity of (120) and (200) reflections increase.
When entering the low-temperature phase the first pair of reflections reveal a sharp increase of intensity, while the second pair behaves in the opposite way.
This behaviour suggests the presence of more than one spin component, e.g. a spin canting, when approaching the second transition, as discussed below.

At this point, we were unable to derive the propagation vector of the low-temperature magnetic structure from the powder data only, which is due to the fact that not all observed magnetic satellites (inspected in the difference patterns by subtracting a higher-temperature background) originated from the low-temperature magnetic phase of \sgo\ (both the monoclinic and cubic modifications of Gd$_2$O$_3$ exhibit a magnetic phase transition between \Tnone\ and \Tntwo\ \cite{Stewart_1979,Child_1967,Moon_1975}).
We therefore relied on further single-crystal experiments which are presented in Sec.~\ref{sec:scnd}.

\begin{figure}[tb]
\includegraphics[width=0.9\columnwidth]{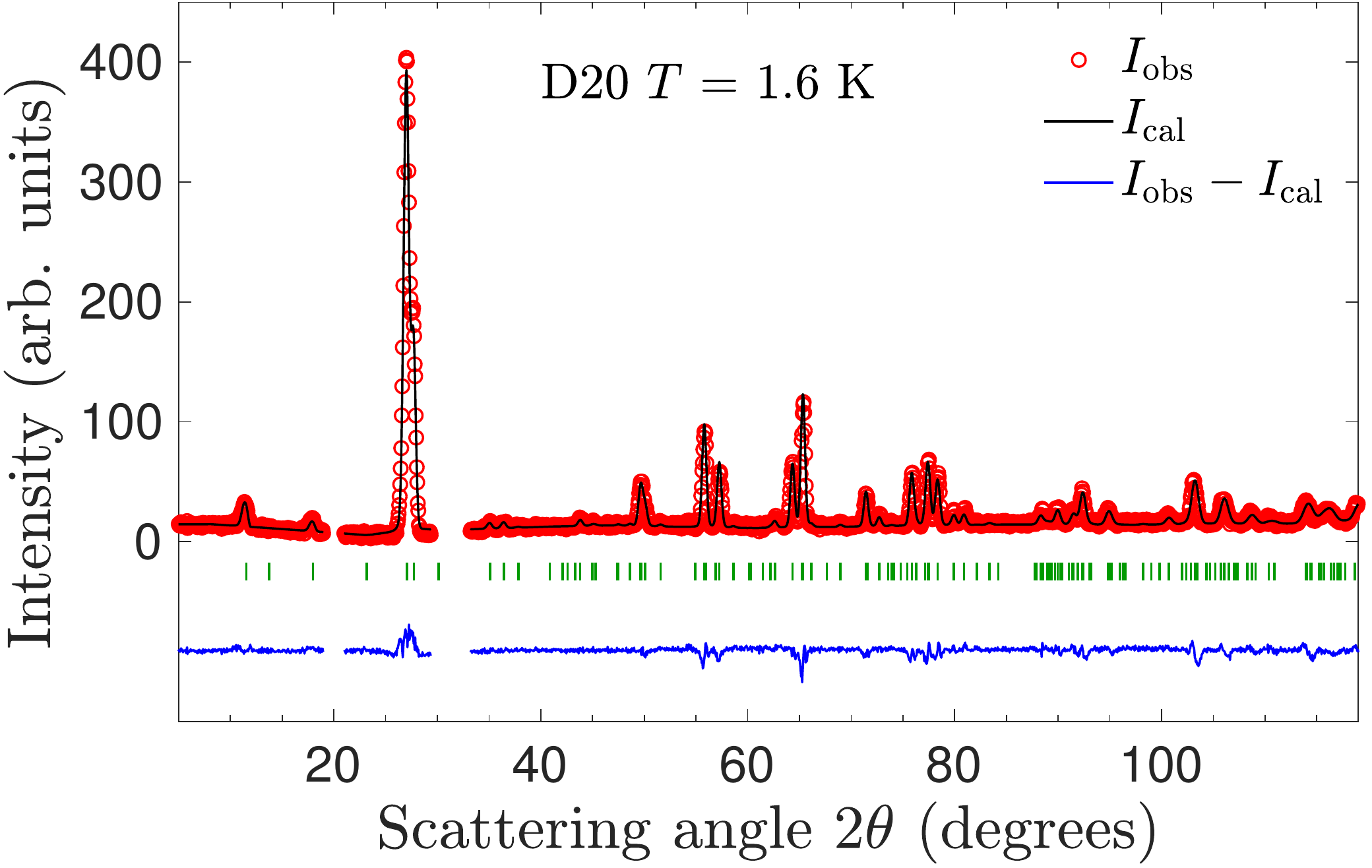}
\caption{Magnetic scattering of \sgo\ at $T=1.6$~K (red dots) extracted by subtracting a non-magnetic background measured at 4~K (D20, $\lambda_{\rm{D20}}$ = 2.41~\AA).
The calculated pattern (black line) corresponds to the $\Gamma_4$ symmetry shown in Table~\ref{tab:irreps1} and agrees very well with the data as evidenced by the difference curve (blue line).
The green markers indicate the positions of the commensurate \afm\ peaks. 
The small impurity peaks at $2\theta$ = 20$^\circ$ and 31$^\circ$ originating from the monoclinic Gd$_2$O$_3$ phase were excluded from the refinement.
The broad dip in the diffraction pattern around $2\theta$ = 30$^\circ$ results from the subtraction of the 4~K signal containing a significant amount of diffuse scattering (as clearly visible in Fig.~\ref{fig:nucpattern}).}  
\label{fig:magpattern}
\end{figure}

In order to derive the magnetic structure between \Tnone\ and \Tntwo\ we have extracted the purely magnetic scattering by analyzing the difference between the patterns at 1.6 and at \mbox{4 K} (nuclear background), which is shown in Fig.~\ref{fig:magpattern}.
Representation analysis was employed using the \textsc{Mag2Pol} program in order to derive magnetic structure models whose symmetry is compatible with the underlying crystal structure and the propagation vector $\mathbf{q}_1$ = (0 0 0).
8 one-dimensional irreducible representations (irreps) are obtained for the space group $Pnam$ (shown in Table~\ref{tab:irreps1}) which were individually tested on the data.
The best agreement was achieved with $\Gamma_4$ yielding a magnetic $R_{\rm F}$ of 10.55 by keeping all previously refined parameters, e.g. the scale factor, atomic positions, etc., fixed.
The only refined magnetic parameters were the $w$ components of $\psi_4$ for each of the Gd sites, which amount to 6.80(3)~$\mu_{\rm B}$ and 4.59(3)~$\mu_{\rm B}$. 
Note that it is not possible to distinguish the two Gd sites in a diffraction experiment as the two possibilities yield the same results within the error bars with similar agreement factors.
But the analysis of crystal fields acting on the \gdi\ ions at the Gd1 and Gd2 sites (see section~\ref{CEF}) indicates that the larger moment should be attributed  to the Gd1 site.
Given that in \seo\ and \sho\ the magnetic order is dominated by a single site, we have verified that this is indeed not the case in \sgo.
A further clear argument for magnetic order on both Gd sites in \sgo\ is the absence of diffuse scattering below \Tntwo\ (not shown), while a strong diffuse signal was visible at the lowest temperatures in both \seo~\cite{Petrenko_2008} and \sho~\cite{Young_2012}. 
The magnetic scattering of \sgo\ at $T=1.6$~K is shown together with the calculated pattern and the difference curve in Fig.~\ref{fig:magpattern}. 

\begin{table*}[tb]
\caption{Basis vectors $\psi_{n}$ of the irreducible representation $\Gamma_{n}$ for the Gd1 and Gd2 sites in \sgo\ for space group \textit{Pnam} and propagation vector $\mathbf{q}_1$ = (0 0 0).}
\label{tab:irreps1}
	\begin{ruledtabular}
		\begin{tabular}{cccccccccc}
			Atom  &  Position   &  $\psi_1$  & $\psi_2$ & $\psi_3$ & $\psi_4$ & $\psi_5$ & $\psi_6$ & $\psi_7$ & $\psi_8$ \\ \hline \\ 
			1  &  $\begin{pmatrix} x \\ y \\ \nicefrac{1}{4} \end{pmatrix}$ &  $\begin{pmatrix}  0  \\  0  \\ w  \end{pmatrix}$ & $\begin{pmatrix}  u  \\  v  \\  0  \end{pmatrix}$ & $\begin{pmatrix}  u  \\  v  \\  0  \end{pmatrix}$ & $\begin{pmatrix}  0  \\  0  \\ w  \end{pmatrix}$ & $\begin{pmatrix}  0  \\  0  \\ w  \end{pmatrix}$ & $\begin{pmatrix}  u  \\  v  \\  0  \end{pmatrix}$ & $\begin{pmatrix}  u  \\  v  \\  0  \end{pmatrix}$ & $\begin{pmatrix}  0  \\  0  \\ w  \end{pmatrix}$ \\ \\ 
			2  &  $\begin{pmatrix} \bar{x} + \nicefrac{1}{2} \\ y + \nicefrac{1}{2} \\ \nicefrac{3}{4}  \end{pmatrix}$ &  $\begin{pmatrix}  0  \\  0  \\ \bar{w}  \end{pmatrix}$ & $\begin{pmatrix}  \bar{u}  \\  v  \\  0  \end{pmatrix}$ & $\begin{pmatrix}  \bar{u}  \\  v  \\  0  \end{pmatrix}$ & $\begin{pmatrix}  0  \\  0  \\ \bar{w}  \end{pmatrix}$ & $\begin{pmatrix}  0  \\  0  \\ w  \end{pmatrix}$  & $\begin{pmatrix}  u  \\  \bar{v}  \\  0  \end{pmatrix}$ & $\begin{pmatrix}  u  \\  \bar{v}  \\  0  \end{pmatrix}$ & $\begin{pmatrix}  0  \\  0  \\ w  \end{pmatrix}$ \\ \\ 
			3  &  $\begin{pmatrix} \bar{x} \\ \bar{y} \\ \nicefrac{3}{4} \end{pmatrix}$ &  $\begin{pmatrix}  0  \\  0  \\  w  \end{pmatrix}$ & $\begin{pmatrix}  \bar{u}  \\  \bar{v}  \\  0  \end{pmatrix}$ & $\begin{pmatrix}  u  \\  v  \\  0  \end{pmatrix}$ & $\begin{pmatrix}  0  \\  0  \\ \bar{w}  \end{pmatrix}$ & $\begin{pmatrix}  0  \\  0  \\ w  \end{pmatrix}$ & $\begin{pmatrix}  \bar{u}  \\  \bar{v}  \\  0  \end{pmatrix}$ & $\begin{pmatrix}  u  \\  v  \\  0  \end{pmatrix}$ & $\begin{pmatrix}  0  \\  0  \\ \bar{w}  \end{pmatrix}$ \\ \\ 
			4  &  $\begin{pmatrix} x+\nicefrac{1}{2} \\ \bar{y}+\nicefrac{1}{2} \\ \nicefrac{1}{4}  \end{pmatrix}$ &  $\begin{pmatrix}  0  \\  0  \\ \bar{w}  \end{pmatrix}$ & $\begin{pmatrix}  u  \\  \bar{v}  \\  0  \end{pmatrix}$ & $\begin{pmatrix}  \bar{u}  \\  v  \\  0  \end{pmatrix}$ & $\begin{pmatrix}  0  \\  0  \\ w  \end{pmatrix}$  & $\begin{pmatrix}  0  \\  0  \\ w  \end{pmatrix}$ & $\begin{pmatrix}  \bar{u}  \\  v  \\  0  \end{pmatrix}$ & $\begin{pmatrix}  u  \\  \bar{v}  \\  0  \end{pmatrix}$ & $\begin{pmatrix}  0  \\  0  \\ \bar{w}  \end{pmatrix}$ \\ 
		\end{tabular}
	\end{ruledtabular}
\end{table*}

\begin{figure}[tb]
\includegraphics[width=0.8\columnwidth]{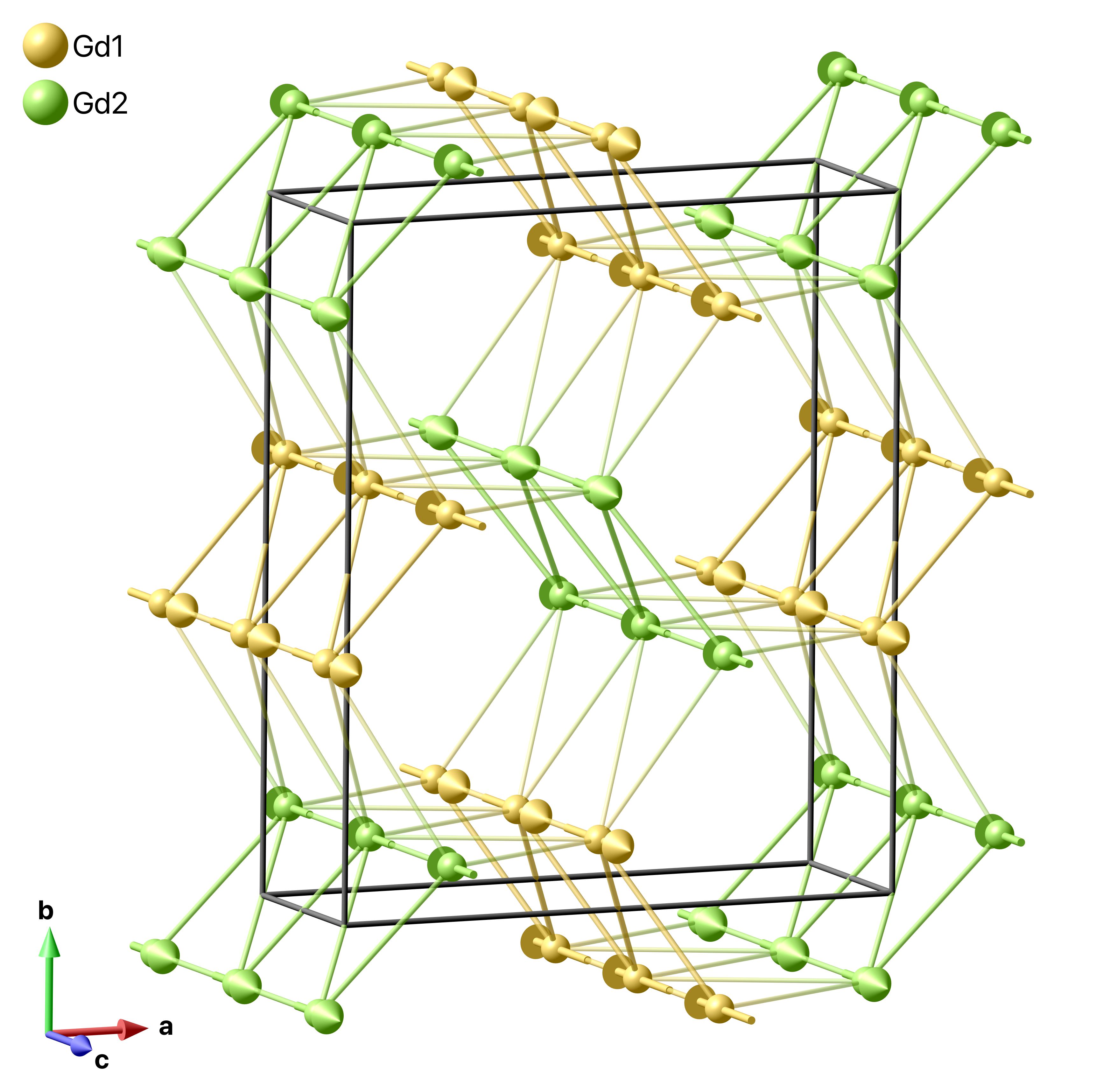}
\caption{Magnetic structure of \sgo\ observed between \Tnone\ and \Tntwo.
The structure is characterized by chains of parallel magnetic moments on the legs of the zig-zag ladders, the moments are parallel or antiparallel to the $c$~axis. The moments on the ladder's rungs are antiparallel to one another.}
\label{fig:magstructure1}
\end{figure}

The resulting magnetic structure consists of chains of parallel spins along the $c$~axis, i.e. the legs of the zig-zag ladders.
Such a leg is antiferromagnetically coupled not only to the second leg of the same Gd site forming the zig-zag ladder (at a distance of 3.62~\AA), but also to the 2 neighboring ladders of the respectively other Gd site (at distances of 3.80~\AA\ and 4.13~\AA).
The magnetic structure is depicted in Fig.~\ref{fig:magstructure1}.

Having established the magnetic structure at $T=1.6$~K we now return to the peculiar temperature dependence of the magnetic intensities below $T=1.45$~K shown in Fig.~\ref{fig:tdep_pow}(a).
Although the variation of intensity for different Bragg reflections - and therefore different projections of the magnetic moments - might indicate the evolution of a secondary spin component besides the main component along the $c$ axis, it can be shown that the magnetic structure (as pictured in Fig.~\ref{fig:magstructure1}) does not qualitatively change.
For that purpose, all diffraction patterns below 3.3~K (excluding the satellite reflections for patterns below \Tntwo) were analyzed using the same model, i.e. a single component $w$ within $\Gamma_4$ symmetry, showing no necessity to include a further spin component.
The resulting temperature dependence of the magnetic moments of the Gd1 and Gd2 sites is shown in Fig.~\ref{fig:tdep_pow}(b).
While both sites show a steady increase when cooling through \Tnone, it is the Gd2 site which shows a stronger increase when approaching \Tntwo\ and a more pronounced drop when entering the low-temperature phase.
The relation to the intensities shown in Fig.~\ref{fig:tdep_pow}(a) becomes clear when looking at the magnetic structure factors.
While the magnetic structure factors of (120) and (200) reflections are proportional to the sum of the Gd1 and Gd2 moments, those of the (010) and (110) reflections are actually proportional to the difference.
Hence, the fact that $\mu$(Gd2) approaches and then departs from the relatively stable value of $\mu$(Gd1) on cooling towards \Tntwo\ and cooling through \Tntwo, respectively, is entirely responsible for the peculiar temperature dependence of the magnetic intensities.

\subsection{Single-crystal neutron diffraction}  \label{sec:scnd}
A nuclear data set consisting of 155 unique reflections was recorded at 10 K in order to fix the nuclear structure parameters and the scale factor for the subsequent magnetic structure analysis.
The refined parameters were the atomic positions, the isotropic temperature factors (constrained to be the same for the same elements), the scale factor and the the extinction coefficients according to an empirical \textsc{Shelx}-like model~\cite{shelx}.
A linear absorption coefficient of $\mu = 1.97$~mm$^{-1}$ was introduced for the wavelength of $\lambda = 2.36$~\AA.
We note that the data collection range was rather limited, it only extends to $\sin(\theta)/\lambda$ = 0.36~\AA$^{-1}$ which results in relatively high values for the refined isotropic temperature factors and its standard deviation [of the order of $B = 3(1)$~\AA$^2$].
The relatively large $B$ value is related to the imperfect absorption correction.
Nevertheless, the atomic positions (shown in Table~\ref{tab:nucsc}) are the same to within error when compared with the powder results and agreement with the observed data is excellent as evidenced by $R_{\rm F} = 3.75$.
We also want to emphasize the importance of an appropriate absorption correction as the refinement on an uncorrected data set yields $R_{\rm F} = 6.79$.

\begin{table}[tb]
\caption{Basis vectors $\psi_{n}$ of the irreducible representation $\Gamma_{n}$ for the Gd1 and Gd2 sites in \sgo\ for space group \textit{Pnam} and propagation vector $\mathbf{q}_2$ = (0 0 0.42).
The phase factor $a$ is given by $\exp(i \pi q_z)$ and originates from the glide plane $n$(0,1/2,1/2) 1/4,$y$,$z$ and the screw axis 2(0,0,1/2) 0,0,$z$, which map atom 1 onto atoms 2 and 3, respectively.}
	\label{tab:irreps2}
	\begin{ruledtabular}
		\begin{tabular}{cccccc}
			Atom  &  Position   &  $\psi_{1}$  & $\psi_2$ & $\psi_3$ & $\psi_4$  \\ \hline \\ 
			1  &  $\begin{pmatrix} x \\ y \\ \nicefrac{1}{4} \end{pmatrix}$ &  $\begin{pmatrix}  u  \\  v  \\  w  \end{pmatrix}$ & $\begin{pmatrix}  u  \\  v  \\  w  \end{pmatrix}$ & $\begin{pmatrix}  u  \\  v  \\  w  \end{pmatrix}$ & $\begin{pmatrix}  u  \\  v  \\ w  \end{pmatrix}$  \\ \\ 
			2  &  $\begin{pmatrix} \bar{x} + \nicefrac{1}{2} \\ y + \nicefrac{1}{2} \\ \nicefrac{3}{4}  \end{pmatrix}$ &  $a\begin{pmatrix}  u  \\  \bar{v}  \\ \bar{w}  \end{pmatrix}$ & $a\begin{pmatrix}  \bar{u}  \\  v  \\  w  \end{pmatrix}$ & $a\begin{pmatrix}  \bar{u}  \\  v  \\  w  \end{pmatrix}$ & $a\begin{pmatrix}  u  \\  \bar{v}  \\ \bar{w}  \end{pmatrix}$  \\ \\ 
			3  &  $\begin{pmatrix} \bar{x} \\ \bar{y} \\ \nicefrac{3}{4} \end{pmatrix}$ &  $a\begin{pmatrix}  \bar{u}  \\  \bar{v}  \\  w  \end{pmatrix}$ & $a\begin{pmatrix}  \bar{u}  \\  \bar{v}  \\  w  \end{pmatrix}$ & $a\begin{pmatrix}  u  \\  v  \\  \bar{w}  \end{pmatrix}$ & $a\begin{pmatrix}  u  \\  v  \\ \bar{w}  \end{pmatrix}$ \\ \\ 
			4  &  $\begin{pmatrix} x+\nicefrac{1}{2} \\ \bar{y}+\nicefrac{1}{2} \\ \nicefrac{1}{4}  \end{pmatrix}$ &  $\begin{pmatrix}  \bar{u}  \\  v  \\ \bar{w}  \end{pmatrix}$ & $\begin{pmatrix}  u  \\  \bar{v}  \\  w  \end{pmatrix}$ & $\begin{pmatrix}  \bar{u}  \\  v  \\  \bar{w}  \end{pmatrix}$ & $\begin{pmatrix}  u  \\  \bar{v}  \\ w  \end{pmatrix}$  \\ 
		\end{tabular}
	\end{ruledtabular}
\end{table}

\begin{table}[tb]
\caption{Structural parameters of \sgo\ within space group \textit{Pnam} refined from the single-crystal experiment on D10. All atoms occupy the Wyckoff position 4$c$ ($x$ $y$ $\nicefrac{1}{4}$).
The lattice parameters were refined from a small set of reflections during the alignment and amount to $a=10.05(3)$~\AA, $b=11.97(4)$~\AA\ and $c=3.458(8)$~\AA.
The extinction parameters are $x_{11}=0.03(2)$, $x_{22}=0.006(4)$, $x_{33}=-0.002(4)$, $x_{12}=0.02(1)$, $x_{23}=-0.04(2)$ and $x_{13}=-0.08(4)$.}
\label{tab:nucsc}
	\begin{ruledtabular}
		\begin{tabular}{ccc}
			Atoms	& $x$		&  $y$		\\ \hline
			Sr		&  0.748(3)	&  0.648(2)	\\ 
			Gd1		&  0.425(4)	&  0.113(2) 	\\ 
			Gd2		&  0.416(4)	&  0.612(2)	\\ 
			O1		&  0.716(3)	&  0.315(3)	\\ 
			O2		&  0.632(4)	&  0.015(3)	\\ 
			O3		&  0.507(3)	&  0.787(3)	\\ 
			O4		&  0.926(3)	&  0.078(3)  
		\end{tabular}
	\end{ruledtabular}
\end{table}

\begin{figure}[tb]
\includegraphics[width=0.9\columnwidth]{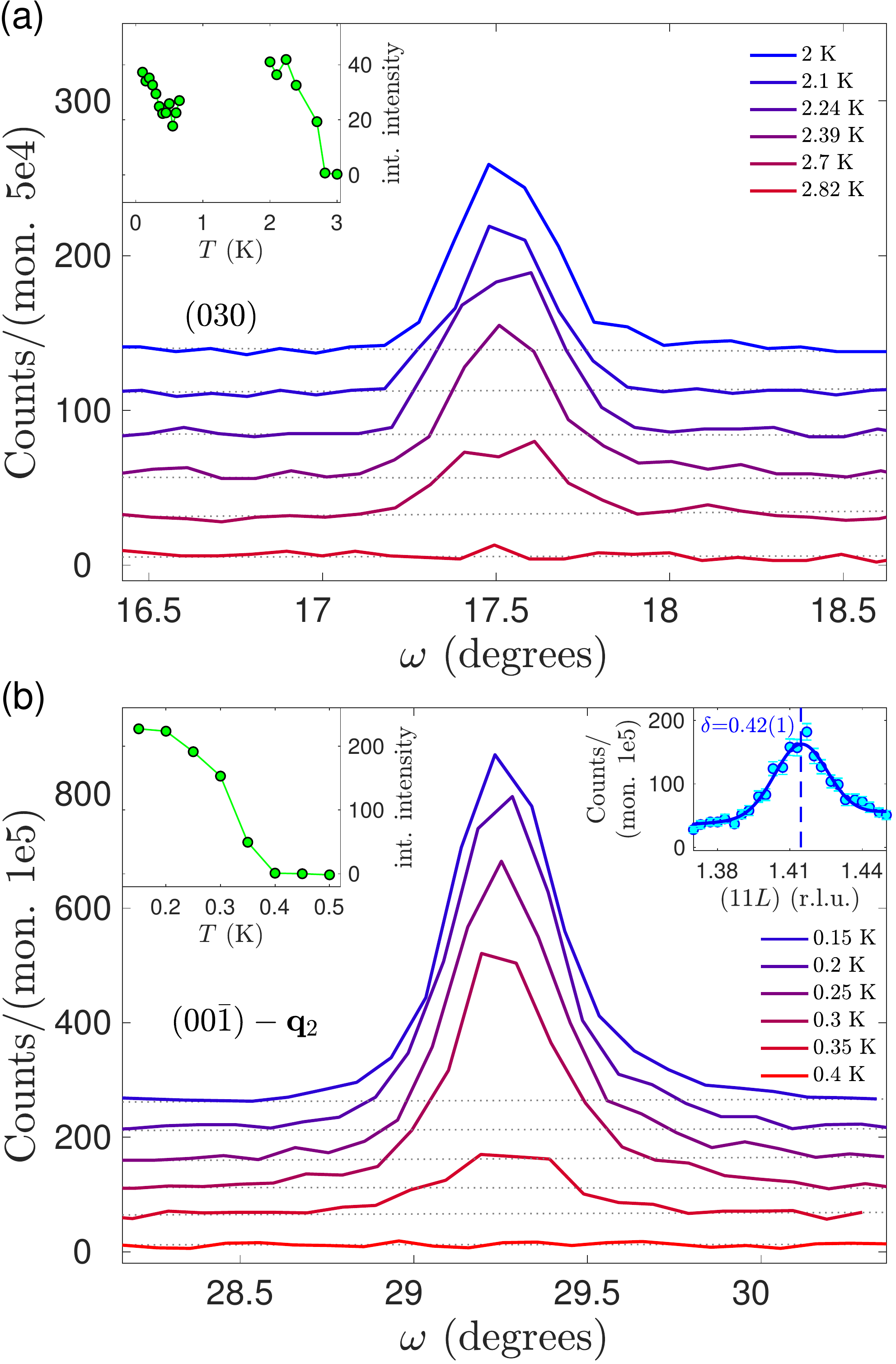}
\caption{(a) Temperature evolution of the $\omega$ scan through the commensurate (030) reflection across the magnetic phase transition at \Tnone.
The integrated intensities shown in the inset reveal an onset of magnetic scattering at $T= 2.8$~K.
Although the $\omega$ scans are not shown for $T<1$~K, the integrated intensities are added to the inset, which forms an overall temperature dependence reminiscent of that of the (010) reflection observed in the powder data (see Fig.~\ref{fig:tdep_pow}).
(b) Temperature evolution of the $\omega$ scan through the incommensurate satellite (00$\bar{1}$) - $\mathbf{q}_2$ reflection across the magnetic phase transition at \Tntwo.
From the integrated intensities shown in the upper left inset the magnetic phase transition temperature seems to be a bit lower than the reported 0.48~K which is probably due to the temperature calibration and homogeneity of the 4-circle dilution cryostat.
The upper right inset shows a reciprocal space scan along the $L$ direction for the (111) + $\mathbf{q}_2$ satellite revealing an incommensurability of $\delta$ = 0.42(1).}
\label{fig:tdep_sc}
\end{figure}

On cooling below \Tnone\ we observe a significant increase in the intensity of the (030) reflection due to the onset of long-range magnetic order [Fig.~\ref{fig:tdep_sc}(a)].
The inset in Fig.~\ref{fig:tdep_sc}(a) shows the integrated intensities as a function of temperature revealing a transition temperature of slightly below 2.8~K, in a good agreement with the published value of \Tnone\ = 2.73~K.
Although for clarity the raw data below 1~K are not shown in the main panel, the integrated intensities are added to the inset showing qualitatively the same temperature dependence as derived for the (010) reflection from the powder data (see Fig.~\ref{fig:tdep_pow}).

The principal magnetic structure was investigated at \mbox{$T$ = 700 mK}, i.e. just above \Tntwo, and at the lowest attainable temperature \mbox{$T$ = 100 mK} using a dataset of 271 unique reflections.
The refinement clearly confirmed the solution derived from the powder measurements: we find the magnetic moments of the Gd1 and Gd2 sites to be 6.4(2)~$\mu_{\rm B}$ and 4.9(2)~$\mu_{\rm B}$ at $T=700$~mK ($R_{\rm F} = 8.72$) and 6.8(2)~$\mu_{\rm B}$ and 4.3(2)~$\mu_{\rm B}$ ($R_{\rm F}=8.20$) at $T=100$~mK, which is in good agreement with the D20 data, although the values resulting from the powder refinement tend to be a bit higher (cf. Fig.~\ref{fig:tdep_pow}).
However, it has to be stressed that even with a 98.4(1)\% $^{160}$Gd enrichment the absorption effects are still quite high and that modeling these effects is rather difficult, especially for a single-crystal sample.
The relatively small discrepancies are therefore a natural consequence of the crystal shape approximation and the thickness estimation of the powder sample, which both directly affect the resulting magnetic moment values.

To derive the low-temperature magnetic structure we performed scans along the high-symmetry directions in reciprocal space at $T = 100$~mK in order to detect any magnetic scattering not present above 500~mK.
We found such satellites at the positions modulated by the propagation vector \mbox{$\mathbf{q}_2$ = (0 0 0.42)} which is shown for the (111) + $\mathbf{q}_2$ reflection in the top right inset of Fig.~\ref{fig:tdep_sc}(b).
The main panel of Fig.~\ref{fig:tdep_sc}(b) shows the rocking curves of the (00$\bar{1}$) - $\mathbf{q}_2$ reflection between 0.4 and 0.15~K confirming the transition temperature \Tntwo.
From the top left inset of Fig.~\ref{fig:tdep_sc}(b) - showing the integrated intensities of the (00$\bar{1}$) - $\mathbf{q}_2$ reflection as a function of temperature - the onset of the low-temperature magnetic phase seems to be located at a slightly lower temperature, which can be explained by a temperature gradient between the sample and the Cernox temperature sensor. 

\begin{figure}[tb]
\includegraphics[width=0.9\columnwidth]{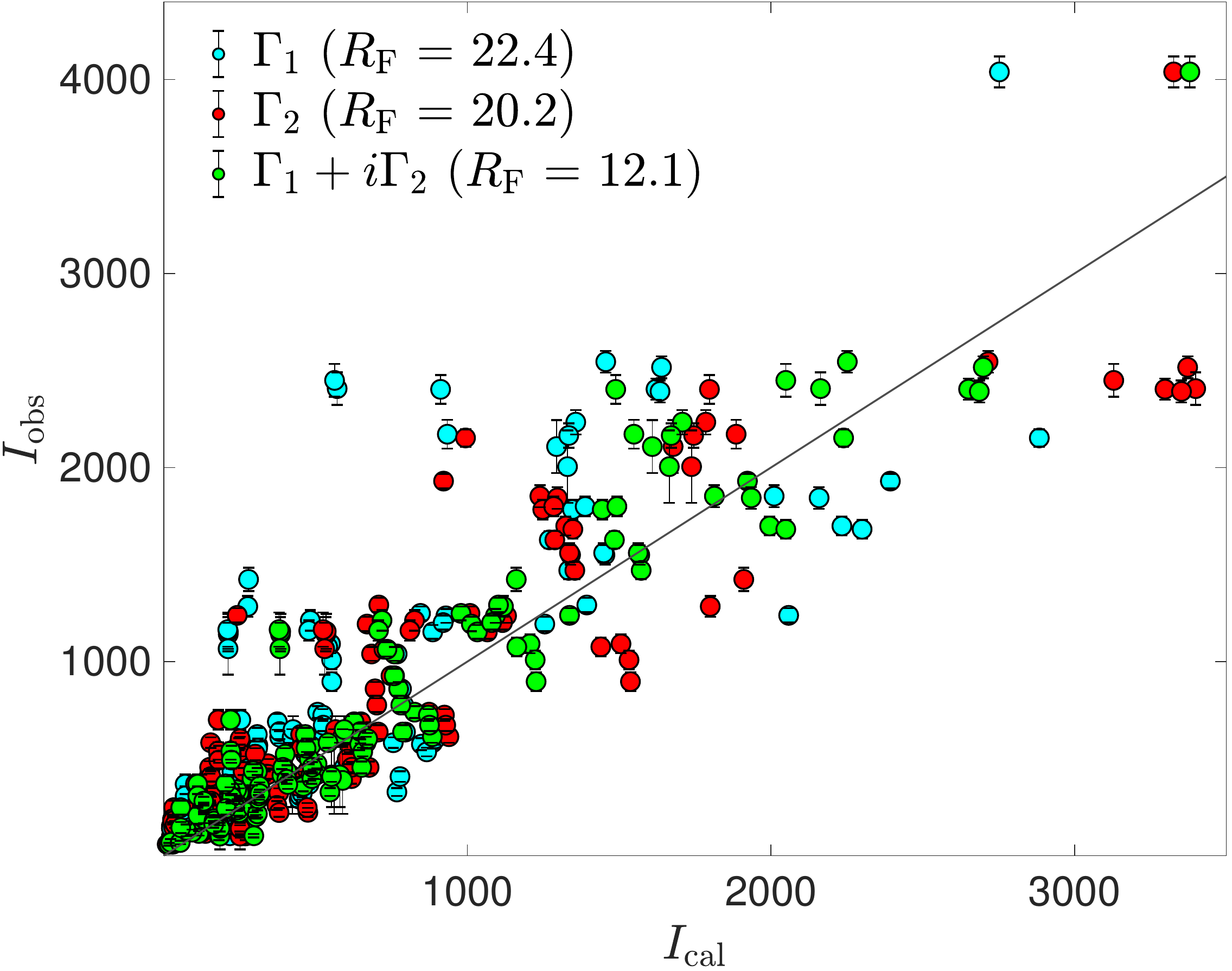}
\caption{Results of the single-crystal refinements using models $\Gamma_1$, $\Gamma_2$ and the mixed representation $\Gamma_1 + i\Gamma_2$ on the satellite intensities observed at 100 mK.}
\label{fig:SCresults}
\end{figure}

As for the high-temperature magnetic phase, representation analysis was employed in order to derive the magnetic structure models.
The calculation for space group $Pnam$ and a propagation vector $\mathbf{q}_2$ = (0 0 0.42) reveals 4 one-dimensional irreducible representations which are listed in Table~\ref{tab:irreps2}.
The 4 models were tested individually on a data set containing 170 unique reflections.
Fair agreement was observed for the models $\Gamma_1$ and $\Gamma_2$, both having $R_{\rm F}$ values of slightly over 20, which is far from a convincing refinement.
For $\Gamma_1$, a strong magnetic moment along the $b$ axis is found for one of the two Gd sites, whereas the $\Gamma_2$ symmetry favors the moment to be lying along the $a$ axis.
With both models not being sufficiently well adapted to describe the experimental data, we decided to mix the basis vectors $\psi_1$ and $\psi_2$ (only the $u$ and $v$ components) in phase quadrature, effectively reducing the magnetic symmetry of the system.
This approach resulted in a reasonably good refinement with an \mbox{$R_{\rm F} = 12.1$}.
Figure~\ref{fig:SCresults} compares the observed and calculated intensities for the models $\Gamma_1$, $\Gamma_2$ and $\Gamma_1 + i\Gamma_2$.

\begin{figure}[tb]
\includegraphics[width=0.8\columnwidth]{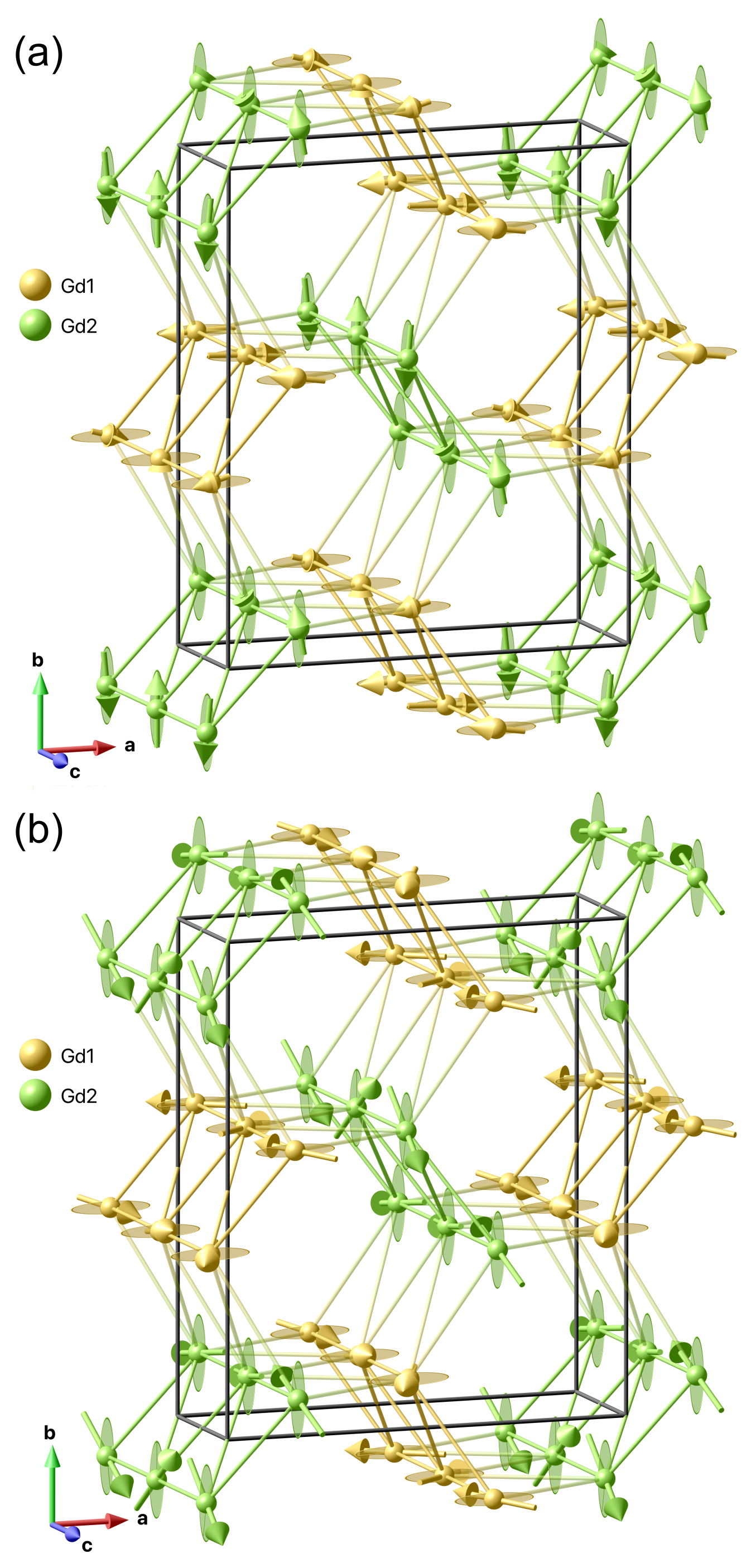}
\caption{Magnetic structure of \sgo\ below \Tntwo.
(a) shows only the incommensurably modulated component along $\mathbf{q}_2$ = (0 0 0.42). This component forms a magnetic structure which can be described as a helix within the $ab$~plane with a strongly elliptic spin-rotation envelope elongated along the $a$~axis for the Gd1 site and along the $b$~axis for the Gd2 site, respectively.
(b) shows the superposition of the commensurate $\mathbf{q}_1$ component and the incommensurate $\mathbf{q}_2$ component [depicted in Fig.~\ref{fig:magstructure1} and (a), respectively] resulting in a conical structure, which - due to the small secondary axis of the envelope - is better described as a fan-like structure.}
\label{fig:magstructure2}
\end{figure}

Interestingly, the moment on the Gd1 site is primarily aligned along the $a$ axis, while the moment on the Gd2 site is mostly parallel to the $b$ axis. At this point it has to be stressed again that our diffraction data are not capable of distinguishing between the two Gd sites, i.e. a similar agreement can be achieved by swapping the sites.
A significant non-zero component perpendicular to the main component within the $\Gamma_1+i\Gamma_2$ symmetry results in a helical modulation where the spin-rotation traces an elongated ellipse. 
The refined complex Fourier coefficients are given by $S(\rm{Gd1})=[4.6(1)\ 0.9(2)i\ 0]$~$\mu_{\rm B}$ and $S(\rm{Gd2})=[0.8(1)\  4.9(2)i\ 0]$~$\mu_{\rm B}$.
The magnetic structure only containing the $\mathbf{q}_2$ component is shown in Fig.~\ref{fig:magstructure2}(a).
By combining the $\mathbf{q}_1$ component which is aligned along the $c$ axis with the helical component of $\mathbf{q}_2$ within the $ab$~plane one obtains a conical structure, which, due to its elongated envelopes, is better described as a fan-like structure.
The ellipses at the atomic positions represent the elongated spin-rotation plane and therefore the local site anisotropy in the $ab$~plane.
The superposition of the two components yields a maximum moment amplitude of 8.2(2)~$\mu_{\rm B}$ and 6.5(2)~$\mu_{\rm B}$ for the Gd1 and Gd2 site, respectively.

\begin{figure}[tb]
\vspace{0.5cm}
\includegraphics[width=0.9\columnwidth]{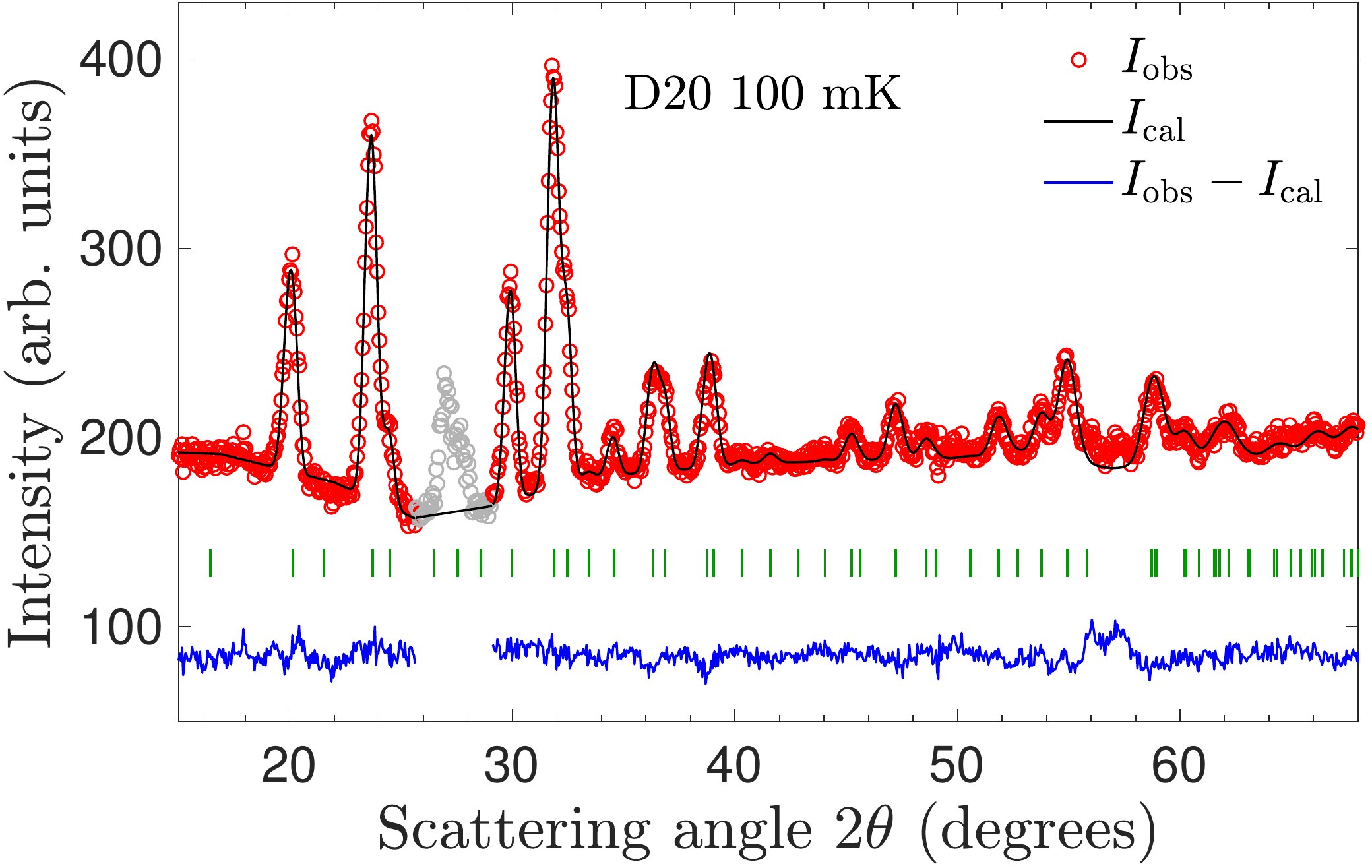}
\caption{Magnetic scattering of the low-temperature phase below \Tntwo\ extracted by taking the difference of diffraction patterns measured at 100~mK and 1~K.
The calculation (black line) based on the model derived from the single-crystal experiment matches the observed data (red dots) very well.
The gray dots were excluded from the refinement as this intensity belongs to the strong commensurate peaks (120) and (200) not being saturated at 1~K. The green markers indicate the positions of the incommensurate \afm\ peaks modulated by $\mathbf{q}_2$.}
\label{fig:magpattern2}
\end{figure}

We have tested the magnetic structure model on our D20 powder data at dilution fridge temperatures.
For that purpose, a difference pattern was calculated by subtracting a 1-K background from the scattering at 100~mK which yields the isolated magnetic satellites.
Note that the difference on commensurate positions (cf. Fig.~\ref{fig:tdep_pow}) could be safely excluded from the refinement for the very strong (120) and (200) reflections (the absolute (120) intensity being approximately 20 times stronger than the strongest satellite reflection) and that the amount on other $\mathbf{q}_1$ positions is marginal in comparison to the strong satellite intensities.
The agreement between the model derived from the single-crystal data with the powder data is excellent as underlined by a magnetic $R_{\rm F} = 8.5$.
By refining the coefficients of the basis vectors of both Gd sites we obtain $S(\rm{Gd1})=[4.44(5)\ 0.69(6)i\ 0]$~$\mu_{\rm B}$ and $S(\rm{Gd2})=[0.47(6)\  4.67(6)i\ 0]$~$\mu_{\rm B}$.
In comparison, the refined magnetic moments are slightly more than $\sigma$ different than the single-crystal values, which, however, amounts to less than 5\% for the main axes of the elliptical envelopes.
The superposition of the magnetic moments modulated by $\mathbf{q}_1$ and $\mathbf{q}_2$ amount to a maximum moment amplitude of 8.12(4)~$\mu_{\rm B}$ and 6.55(5)~$\mu_{\rm B}$ for Gd1 and Gd2, respectively, in excellent agreement with the single-crystal results.
At this point we would like to reiterate that the refined moment size is strongly affected by the applied absorption correction, for which magnetic moments higher than the theoretical value is neither an unusual nor a contradicting result.
We recall that the calculation of the absorption cross section is based on the chemical analysis of the Gd$_2$O$_3$ starting material.
Any uncertainty - especially for the highly absorbing Gd isotopes - can have a big impact on the linear absorption coefficient.
The refined value of the incommensurability is $\delta$ = 0.410(1).
Figure~\ref{fig:magpattern2} shows the observed pattern, the calculated pattern and the difference curve.

\subsection{Modeling of the magnetic properties of \sgo} 
\label{CEF}
\begin{table}[tb] 
\caption{The energies of the CF sublevels of the $^8$S$_{7/2}$ multiplet and the diagonal components of $g$-tensors (in the crystallographic frame) of \gdi\ ions at Gd1 and Gd2 sites corresponding to spin-Hamiltonian $H_{\rm SH}$ and single ion Hamiltonian $H_s$ (in parenthesis).}
\begin{ruledtabular}
	\begin{tabular}{llllllll}
	 \multicolumn{2}{c}{$E~({\rm K})$} & \multicolumn{2}{c}{$g_{aa}$} & \multicolumn{2}{c}{$g_{bb}$} & \multicolumn{2}{c}{$g_{cc}$} \\	\hline
	 \multicolumn{8}{c}{Gd1} \\		\hline
	 0		& (0)		& 0		& (0.09)	& 0		& (0.09)	& 13.98	& (13.54)	\\
	 0.28		& (0.31)	& 0.44	& (2.70)	& 0.46	& (2.33)	& 10.01	& (8.60)	\\
	 0.49		& (0.53)	& 1.76	& (5.58)	& 0.94	& (7.40)	& 5.89	& (3.31)	\\
	 0.62		& (0.80)	& 7.40	& (6.00)	& 8.60	& (12.45)	& 1.91	& (0.27)	\\	\hline
	 \multicolumn{8}{c}{Gd2}  \\	\hline
	 0		& (0)		& 2.78	& (1.52)	& 13.58	& (13.5)	& 0		& (0.56)	\\
	 0.92		& (0.55)	& 1.43	& (3.41)	& 9.70	& (7.94)	& 0.54	& (4.19)	\\
	 1.51		& (1.01)	& 5.0	4	& (8.67)	& 4.97	& (2.51)	& 5.21	& (2.92)	\\
	 1.94		& (1.64)	& 12.44	& (13.70)	& 2.3	1	& (1.25)	& 2.32	& (0.31)
	 \end{tabular}
\end{ruledtabular}
\label{tab:CEF_I}
\end{table}

In order to assign the values of the magnetic moments of the \gdi\ ions measured in the magnetically ordered phases to a particular crystallographic site and also to extract information about the magnetic interactions in \sgo\ we first consider the spectral and magnetic properties of the non-interacting \gdi\ ions positioned at the Gd1 and Gd2 sites as well as the crystal fields (CF) affecting them.
A single-ion Hamiltonian, $\cal H_s=H_{\rm FI}+H_{\rm CF}$, contains only two terms, a free ion ($\cal H_{\rm FI}$) and the CF interaction ($\cal H_{\rm CF}$).
The free ion Hamiltonian operating in the total space of 3432 states of the electronic 4$f^7$ configuration is written in the standard form~\cite{Carnall_1989}.
For the initial simulations, we used the set of CF parameters determined previously for the Er$^{3+}$ ions (ground state configuration 4$f^{11}$) at the Er1 and Er2 sites in \seo\ in Ref.~\cite{Malkin_2015}.
Such a choice is permissible given a typically monotonic variation of the CF parameters along the lanthanide series in the isostructural compounds.
We obtained the energies of the CF sublevels (Kramers doublets) and the corresponding wavefunctions for all electronic multiplets of the \gdi\ ions at Gd1 and Gd2 sites by numerical diagonalization of the single-ion Hamiltonian.

On the other hand, the electron paramagnetic resonance spectra of an isolated \gdi\ impurity in the isostructural diluted paramagnet SrY$_2$O$_4$:Gd (0.5 at.\%) were successfully described by the effective spin-Hamiltonian, $\cal H_{\rm SH}$, operating in the truncated basis of states of the $^8$S$_{7/2}$ multiplet~\cite{Gabbasov_2019}.
The sets of parameters in the spin-Hamiltonians $\cal H_{\rm SH}$ for \gdi\ ions substituting for Y$^{3+}$ ions in two pairs of magnetically nonequivalent Y1 or Y2 sites in the unit cell were found from the analysis of the angular dependencies of the resonant magnetic fields~\cite{Gabbasov_2019}.
The two approaches, based on the CF Hamiltonians assigned to specific positions of the rare earth ions in the crystal lattice and the spin-Hamiltonians introduced in Ref.~\cite{Gabbasov_2019}, bring about qualitatively similar energies of the four CF sublevels of the ground state multiplet $^8$S$_{7/2}$ (see Table~\ref{tab:CEF_I}) and the corresponding $g$-factors.
Thus, it is possible to identify different spin-Hamiltonians with particular Gd1 and Gd2 sites in agreement with the notation used previously in Ref.~\cite{Gabbasov_2019}.
A relatively large total splitting of about 2~K of the ground multiplet correlates well with a larger deformation of an oxygen octahedron at the Gd2 sites.
A strong anisotropy of the $g$-factors presented in Table~\ref{tab:CEF_I} begins to play a role at low temperatures only when the energies of the thermal excitations become comparable to the zero-field splitting of the ground state multiplet.

\begin{figure}[tb]
\includegraphics[width=0.9\columnwidth]{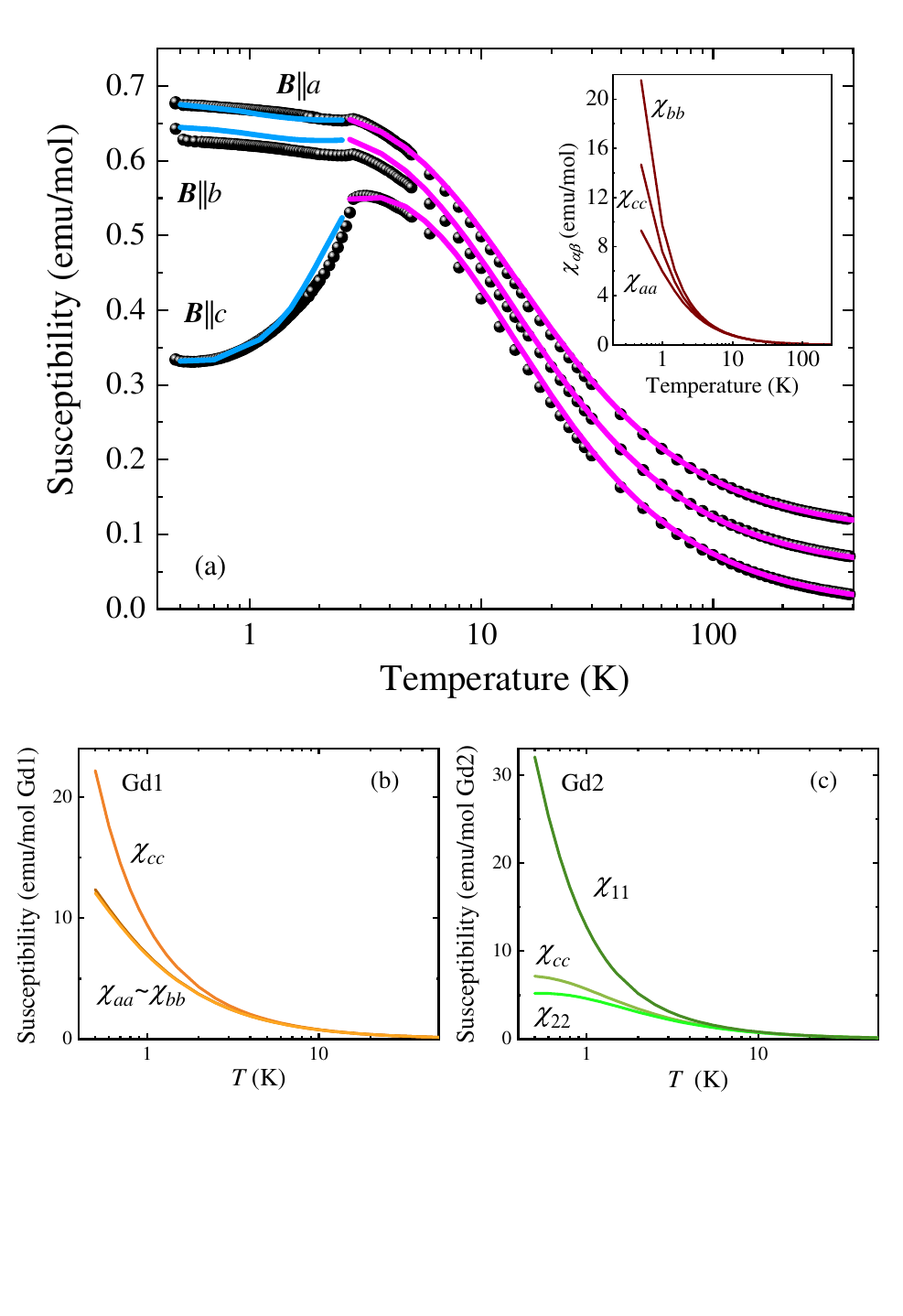}
\vspace{-20mm}
\caption{
(a) Temperature dependence of the measured (symbols)~\cite{Young_2014} and calculated (solid lines) dc-susceptibilities of \sgo,  $B=0.01$~T. 
	The curves are consecutively offset by 0.05~emu/mole for clarity.
	Different colors for the theoretical curves emphasize different approaches taken for calculations at $T_{\rm N2}<T<T_{\rm N1}$ and $T_{\rm N1}<T$, see main text.
	Inset: calculated single ion susceptibilities averaged over the Gd1 and Gd2 sites.
	(b), (c) The susceptibilities separately calculated for the Gd1 and Gd2 sites, respectively.}
\label{fig:chi}
\end{figure} 

The components  of the single-ion magnetic dc~susceptibility tensors of the \gdi\ ions at the Gd1 and Gd2 sites for temperatures in the range from 0.5 to 250~K were calculated by making use of the corresponding spin-Hamiltonians, the results averaged over the two equally populated Gd1 and Gd2 sites are presented in Fig.~\ref{fig:chi}(a) (inset) and the temperature dependencies of the main values of the susceptibility tensors ($\chi_{11}$, $\chi_{22}$, and $\chi_{cc}$) at the Gd1 and Gd2 sites are shown in Figs.~\ref{fig:chi}(b) and \ref{fig:chi}(c). 
One of the main axes of the susceptibility tensors coincides with the crystallographic $c$~axis. 
In the $ab$~plane, the susceptibility tensor at the Gd1 sites is practically isotropic, but there is a remarkable anisotropy at the Gd2 sites at low temperatures where the main axis corresponding to the largest value of the susceptibility, $\chi_{11}$, is tilted from the $b$~axis by approximately $\pm 12^\circ$ at magnetically non-equivalent sites.
Thus, the single-ion magneto-crystalline anisotropy points to a possible easy-axis type of ordering of the magnetic moments at the Gd1 sites along the $c$~axis and a more complex, in particular, canted type of ordering of the magnetic moments at the Gd2 sites in the $ab$-plane.

Next, we determined the values of exchange integrals in isotropic exchange interactions between the \gdi\ ions from modeling the dc-susceptibilities of the concentrated \sgo\ magnet measured in Ref.~\cite{Young_2014}.
Simulations were performed in the framework of the self-consistent four-particle cluster model that was derived in the study of the isostructural erbium oxide \seo~\cite{Malkin_2015}.
The Hamiltonian $\cal H_{\rm Cl}$ of a cluster in a zig-zag chain [see Fig.~\ref{fig:structure}(c)] operating in the space of $8^4=4096$ states belonging to ground multiplets of four \gdi\ ions contains a sum of spin-Hamiltonians corresponding to the Gd1 and Gd2 sites, magnetic dipole and exchange interactions between the first and second neighbors in the zig-zag chain (along rungs and legs, respectively, of a ladder formed by a chain propagating along the $c$~axis), Zeeman interactions of all four ions with the external magnetic field $\mathbf B$ and of the outer ions in a cluster (labeled in Fig.~\ref{fig:structure}(c) with I and IV) with the auxiliary magnetic field $\Delta \mathbf B$ determined from a condition of equal susceptibilities of all ions in a selected cluster.
Interactions between the selected cluster and other clusters in all the zig-zag chains were taken into account by making use of the mean-field approximation.
\begin{widetext}
	\begin{equation}
		\begin{aligned}
		{\cal H}_{\rm Cl}  = \sum_{j={\rm I,IV}}{\cal H}_{{\rm SH},j} &
		+ \mathbf m_{\rm I} \bm J_r \mathbf m_{\rm II} + \mathbf m_{\rm II} \bm J'_r \mathbf m_{\rm III} + \mathbf m_{\rm III} \bm J_r \mathbf m_{\rm IV}
		+ \mathbf m_{\rm I} \bm J_c \mathbf m_{\rm III} + \mathbf m_{\rm II} \bm J_c \mathbf m_{\rm IV} \\
	& - (\mathbf m_{\rm I}+\mathbf m_{\rm IV})\mathbf B_{\rm loc,o} - (\mathbf m_{\rm II}+\mathbf m_{\rm III})\mathbf B_{\rm loc,i}.
		\end{aligned}
		\label{eq:ham}
	\end{equation}
\end{widetext} 
Here, ${\mathbf m}_j = g \mu_{\rm B} {\mathbf S}_j$ ($g=1.994$~\cite{Gabbasov_2019}, $S=7/2$) and ${\mathbf S}_j$ are the magnetic and spin moments, respectively, of the corresponding \gdi\ ion, $\mathbf B_{\rm loc,o}$ and $\mathbf B_{\rm loc,i}$ are the local magnetic fields affecting the outer and inner ions in a cluster, respectively, the three-dimensional matrices
$\bm J_r=\bm J_{{\rm dd},r}+\bm J_{{\rm ex},r}$, $\bm J'_r=\bm J'_{{\rm dd},r}+\bm J'_{{\rm ex},r}$, and $\bm J_c=\bm J_{{\rm dd},c}+\bm J_{{\rm ex},c}$ determine magnetic interactions between the \gdi\ ions and involve contributions from dipole-dipole and isotropic exchange interactions.
The explicit expressions for the local magnetic fields were presented in Ref.~\cite{Malkin_2015}.
Note, the magnetic dipole interactions were considered exactly using the Ewald method in computations of the corresponding lattice sums.
The results of the fitting of susceptibilities of \sgo\ in the paramagnetic phase $(T>T_{\rm N1})$ are shown in Fig.~\ref{fig:chi}(a).
Table~\ref{tab:CEF_II} contains parameters of the exchange interactions used in the final computations which are compared with literature data \cite{Hasan_2017}  and parameters of the dipolar interactions calculated by making use of the structural parameters from Table~\ref{tab:nucsc}.

\begin{table}[tb] 
\caption{Parameters of the interactions between the \gdi\ ions at the Gd1 and Gd2 sites in \sgo\ (in units of $10^{-3} {\rm K}/\mu^2_{\rm B}$) in the crystallographic frame.}
\begin{ruledtabular}
	\begin{tabular}{lccccccccc}
		Bond						& Length	& $J_{\rm ex}$		& $J_{\rm ex}$	&  \multicolumn{6}{c}{$J_{{\rm dd},\alpha \beta}$}	\\
								& (\AA)		& \cite{Hasan_2017}	&			& $aa$	& $bb$	& $cc$	& $ab$	& $ac$		& $bc$	\\	\hline
		Gd1-Gd1\footnotemark[1]		& 3.458		& 19.4			& 51.8		& 15.0	& 15.0	& -30.0	& 0		& 0			& 0		\\
		Gd2-Gd2\footnotemark[1]		& 3.458		& 65.9			& 51.8		& 15.0	& 15.0	& -30.0	& 0		& 0			& 0		\\
		Gd1-Gd1\footnotemark[2]		& 3.547		& 58.2			& 131		& 6.3		& -10.4	& 4.0		& 13.5	& $\mp$8.6	& $\pm$15.5	\\
		Gd2-Gd2\footnotemark[2]		& 3.610		& 42.6			& 137		& 4.6		& -8.6	& 4.2		& 13.8	& $\mp$8.9	& $\pm$14.1	\\
		Gd1-Gd2					& 3.839		& -19.4			& 62.0		\\
		Gd1-Gd2					& 4.047		& 35.0			& 62.0
	\end{tabular}
\end{ruledtabular}
\footnotetext[1]{The corresponding parameters refer to the matrices $\bm J_c$.}
\footnotetext[2]{Parameters of these bonds refer to the matrices $\bm J_r$ (upper signs) and  $\bm J'_r$ (lower signs).}
\label{tab:CEF_II}
\end{table}

It is seen in Fig.~\ref{fig:chi} that the responses of \sgo\ to an external magnetic field are strongly suppressed at low temperatures (by more than an order of magnitude) as compared to those for a system of noninteracting \gdi\ ions in the same crystal fields.
As a consequence of strong \afm\ interactions within the gadolinium triangles in the zig-zag chains, \sgo\ is a strongly frustrated magnet.
However, as follows from magnetometry~\cite{Young_2014} and neutron diffraction data presented above, with decreasing temperature the geometrical frustration is overcome by long-range ferromagnetic dipolar interactions along the legs of the ladders, and \sgo\ undergoes a phase transition accompanied by lining up of magnetic moments along the legs but with different directions in the neighbor legs because of strong exchange and dipolar \afm\ interactions along the rungs of the ladders.

The transition temperature \Tnone\ was estimated from a condition of a singular response of \sgo\ on the staggered magnetic field having opposite directions along the neighbor legs.
The corresponding ``\afm '' susceptibility $\chi_{{\rm af},cc}(T)$ was calculated by making use of exchange integrals from Table~\ref{tab:CEF_II} and the cluster model.
The equation $\chi^{-1}_{{\rm af},cc}(T)=0$ has a solution at $T=3.25$~K.
This value is overestimated because of neglecting thermal fluctuations in the framework of the mean-field approximation, but it is close to the experimental value of 2.73~K.

Further, we calculate the self-consistent spontaneous magnetic moments of the \gdi\ ions at the Gd1 and Gd2 sites considering four-particle clusters in the zig-zag chains embedded into the crystal magnetically ordered in accordance with the magnetic structure described above.
In particular, at $T$ = 0.7~K, we obtained absolute values of 6.5 and 4.3$\mu_{\rm B}$ for magnetic moments along the $c$~axis at the Gd1 and Gd2 sites, respectively, which agree qualitatively with the neutron diffraction data.
The calculated temperature dependencies of the spontaneous moments for $T>0.5$~K can be approximated by the fractional power function $(1-T/T_{\rm N1})^{0.4}$.

Finally, we calculate the temperature dependencies of the dc~susceptibilities in the magnetically ordered phase for temperatures $T_{\rm N2}<T<T_{\rm N1}$ using the cluster Hamiltonian supplemented by the energies of the exchange and dipolar interactions between the ions in a cluster and surrounding ions with temperature dependent spontaneous magnetic moments.
The results of the calculations presented in Fig.~\ref{fig:chi} agree satisfactorily with the experimental data~\cite{Young_2014}.
Thus, we believe that the obtained set of parameters of the constructed microscopic model presented in Table~\ref{tab:CEF_II} is physically meaningful.

\section{Discussion}		\label{sec:disc}
The results presented above testify to the highly complex magnetic arrangements in \sgo.
Understanding of the mechanisms which govern the long-range magnetic ordering and the selection of a particular direction for the magnetic moments as well as the anisotropy of their correlation functions in the multi-sublattice gadolinium compounds remains challenging.

In the case of the isotropic exchange interactions, the magnetic anisotropy is induced by the CF interaction (single-ion anisotropy) and the long-range dipolar interactions.
However, the dipolar interactions also depend on the single-ion physics through the $g$-factors of the CF sublevels.
In \sgo, the lowest Kramers doublets in the ground state multiplet $^8$S$_{7/2}$ have a nearly Ising-type magnetic anisotropy with approximately the same values of the $g$-factors but along orthogonal directions (along and normal to the zig-zag chains) at the Gd1 and Gd2 sites, as can be seen in Table~\ref{tab:CEF_I}.
Despite a rather large magnetic susceptibility of the Gd2 site along the $b$~axis compared to the susceptibility along the $c$~axis of the Gd1 site (note, the susceptibilities involve contributions from all CF sublevels mixed in the external magnetic field), the spontaneous magnetic moments on both sites are parallel to the $c$~axis in the phase between \Tnone\ and \Tntwo.
The observed magnetic structure of a simple Neel type in the zig-zag chains with the propagation vector $\mathbf{q}_1$ = (0 0 0) appears likely due to the dominant role of the strong dipolar interaction which forces parallel alignment of the moments in the chains.
This conjecture agrees well with the results of the Monte-Carlo simulations in Ref.~\cite{Hasan_2017}.
The model used in the simulations consisted of the classical Heisenberg spins coupled by the exchange and the dipolar interactions whereas the spin-orbit interactions (and crystal-field effects) were presumed to be small.
It was found that for a system to demonstrate a double phase transition in zero field, both the exchange and dipolar terms need to be included, otherwise the second, lower-temperature transition was absent.

A more puzzling question concerns the origin of the incommensurate state at the lowest temperature and the complex temperature evolution of the ground state of \sgo\ in general.
One could suppose that the additional components of the magnetic moments stabilized along the $b$~axis at the Gd2 sites appear due to the increasing population of the ground state doublet with deceasing temperature.
The origin of the components of the magnetic moments along the $a$~axis at the Gd1 sites has only a rather qualitative explanation.
As was mentioned above, the susceptibility tensor is practically isotropic in the $ab$-plane at the Gd1 sites [see Fig.~\ref{fig:chi}(b)], but a weak anisotropy arises at temperatures below 1~K.
In particular, the calculated magnetic moments at $T=0.4$~K in the magnetic field $B=0.3$~T along the $a$ and $b$~axis equal 5.27 and 5.23~$\mu_{\rm B}$, respectively.
A marginally larger value for the $a$~axis hints at a more preferable orientation along this direction, but a detailed simulation of the local magnetic fields in the incommensurate phase goes way beyond the scope of this work.

It has to be noted that the previous bulk properties measurements~\cite{Young_2014} were mostly focused on the intermediate temperature phase, $T_{\rm N2}<T<T_{\rm N1}$, while a detailed study of the low-temperature phase, $T<T_{\rm N2}$, was  limited by the experimental capabilities.
Given the incommensurate nature of the low-temperature phase established in this study, an extension of the bulk properties measurements to temperatures well below $T_{\rm N2}$ might offer new insight into an intriguing magnetic state.

\section{Conclusions}
We have presented an extensive neutron diffraction study using both powder and single-crystal samples of \sgo\ containing isotopically enriched $^{160}$Gd in order to decrease absorption effects and render the investigation of the low-temperature magnetic structures possible.
Apart from being the only member of the \slo\ family which reveals two magnetic phase transitions we observe further properties which give the Gd compound a unique role within these geometrically frustrated systems.
In fact, all three orthorhombic directions are present in the local anisotropy of the Gd1 and Gd2 sites.
Incontestably, the easy direction of the system is the one along the chains parallel to the $c$~axis as evidenced by the collinear magnetic structure which is stabilized below \Tnone. 
Nevertheless, at the unique second phase transition \Tntwo\ the fundamentally different behavior between the two sites - emerging from different crystal-field surroundings as a consequence of different Gd-O octahedra distortions - is manifested in the magnetic structure.
Each site develops an additional helically modulated component perpendicular to the collinear component.
The envelope which the spin traces upon translation along $\mathbf{q}_2$ is strongly elongated along the $a$ or the $b$ axis for Gd1 and Gd2, respectively, resulting in a superposed spin configuration which can be described as fan-like. 

An interesting aspect of this study is that the reported results might not be of importance from solely fundamental research viewpoint, as there are potentially two avenues for exploiting \sgo\ in pure and lightly doped forms in applications, the luminescence~\cite{Singh_2015,Singh_2016,Singh_2017,Sun_2018}
and the low-temperature magnetocalorics~\cite{Jiang_2018,Palacios_2022}.

\section*{ACKNOWLEDGMENTS}
We would also like to acknowledge the expertise and dedication of the low-temperature group at the Institut Laue-Langevin.
We are grateful to M.R. Lees for careful reading the manuscript and helpful suggestions.
The work at the University of Warwick was supported by EPSRC through grants EP/M028771/1 and EP/T005963/1.
B.Z.M. acknowledges the support by the Russian Science Foundation, Project No. 19-12-00244.
		
\bibliography{SrLn2O4_all}
\end{document}